\documentclass[%
 reprint,
superscriptaddress,
nofootinbib,
 amsmath,amssymb,
 aps,
]{revtex4-1}

\usepackage{graphicx}
\usepackage{subfigure}% Include figure files

\usepackage[justification=raggedright]{caption}
\usepackage{tabularx}
\usepackage[colorlinks=true,linkcolor=blue,
     urlcolor = blue,
     citecolor=blue,]{hyperref}
\usepackage{enumerate}
\usepackage{mathtools}
\usepackage{tensor}
\usepackage{braket}
\usepackage{wasysym}
\usepackage{mathrsfs}

\usepackage{appendix}
\usepackage{dcolumn}% Align table columns on decimal point
\usepackage{bm}%

   \newcommand{\bmt}{\bm{\theta}}  
\newcommand{\bmxi}{\bm{\xi}}       
    
\newcommand{\bmu}{\bm{u}}     \newcommand{\bmv}{\bm{v}}  
 
\newcommand{\bmg}{\bm{g}}    \newcommand{\bmh}{\bm{h}}
\newcommand{\sky}{\hat{\bm{n}}}  % 天空方向单位向量
\newcommand{\rmd}{{\rm d}}       % 轻量级微分符号
\newcommand{\calz}{\mathcal{Z}}

\newcommand{\cala}{\mathcal{A}}

\newcommand{\Referee}[1]{#1}
\allowdisplaybreaks

\begin{document}

% \preprint{APS/123-QED}

%---------------------------------------------------------------------
\title{Anatomy of parameter-estimation biases in overlapping gravitational-wave
signals: detector network}
% Force line breaks with \\

\author{Ziming Wang}
\affiliation{Department of Astronomy, School of Physics, Peking University,
Beijing 100871, China}
\affiliation{Kavli Institute for Astronomy and Astrophysics, Peking University,
Beijing 100871, China}

\author{Dicong Liang}
\affiliation{Department of Mathematics and Physics, School of Biomedical Engineering, 
Southern Medical University, Guangzhou, 510515, China}

\author{Lijing Shao}
\email[Corresponding author: ]{lshao@pku.edu.cn}
\affiliation{Kavli Institute for Astronomy and Astrophysics, Peking University,
Beijing 100871, China}
\affiliation{National Astronomical Observatories, Chinese Academy of Sciences,
Beijing 100012, China}
%---------------------------------------------------------------------

%---------------------------------------------------------------------
\begin{abstract}
With the significantly improved sensitivity and a wider frequency band, the
next-generation gravitational-wave (GW) detectors are anticipated to detect
$\sim 10^5$ GW signals per year with durations from hours to days, leading to
inevitable signal overlaps in the data stream. \Referee{While a direct fitting for all signals
may be challenging, extracting only one signal will be biased by its overlap
with other signals.} From this perspective, understanding how the biases arise
from the overlapping and their dependence on the signal parameters is crucial
for developing effective algorithms. In this work, we extend the anatomy of
biases in single-detector cases~\cite{Wang:2023ldq} to a detector network.
\Referee{Specifically, we examine how the biases of the chirp mass, symmetric mass ratio, luminosity distance, and coalescence time depend on the source's sky position and orientation, as well as on the coalescence time and phase.} We propose a
new quantity, named the bias integral, as a useful tool, and establish
relationship between the biases in a single detector and that in the entire
network, with explicit dependence on extrinsic parameters.  Using a 3-detector
network as an example, we further explore the potential of a network to suppress
biases due to the detectors' different locations and orientations.  We find that
location generally has a smaller effect than orientation, and becomes
significant only when the time separation between signals is below sub-seconds.
Through a population-level simulation over the extrinsic parameters, we find
that nearly half of overlapping signals will lead to larger biases in the
network compared to a single detector, highlighting the need to cope with
overlapping biases in a detector network.
\end{abstract}
%---------------------------------------------------------------------

\maketitle

%---------------------------------------------------------------------
\section{Introduction}
%---------------------------------------------------------------------

The direct detection of gravitational waves (GWs)~\cite{Abbott:2016blz} has
opened a new avenue for exploring the Universe. Since the first detection in
2015, the LIGO-Virgo-KAGRA collaboration has reported more than 200 GW events
from compact binary coalescences (CBCs)~\cite{LIGOScientific:2018mvr,
LIGOScientific:2020ibl, LIGOScientific:2021usb, LIGOScientific:2021djp,
LIGOScientific:2025hdt, LIGOScientific:2025slb}, which greatly enrich our
knowledge of astrophysics~\cite{LIGOScientific:2020kqk, KAGRA:2021duu,
LIGOScientific:2025pvj}, fundamental physics~\cite{LIGOScientific:2016lio, 
LIGOScientific:2018dkp, LIGOScientific:2019fpa, LIGOScientific:2020tif,
LIGOScientific:2021sio}, nuclear physics~\cite{LIGOScientific:2018cki,
LIGOScientific:2018hze} and cosmology~\cite{LIGOScientific:2017adf}.  The number
of detected events is rapidly increasing with a rate of several events per week
currently~\cite{LIGOScientific:2020ibl, LIGOScientific:2021djp, LIGO2024}.  The
next-generation (XG) ground-based GW detectors, such as the Cosmic
Explorer~\cite{Reitze:2019iox, Reitze:2019dyk} and the Einstein
Telescope~\citep{Punturo:2010zz, Hild:2010id, Sathyaprakash:2012jk,
Abac:2025saz}, are under development, expected to have about one order of
magnitude higher sensitivity and a wider frequency band than the current LIGO
detectors~\cite{LIGOScientific:2016wof, KAGRA:2013rdx, Kalogera:2021bya}.  In
the future GW observations, it is expected to detect ${\cal O}(10^5)$ CBC events
per year with effective durations around hours to days~\cite{Maggiore:2019uih,
Kalogera:2021bya, Samajdar:2021egv, Himemoto:2021ukb, Hu:2022bji, Wu:2022pyg,
Johnson:2024foj}.  Under such circumstances, there will be cases that multiple
GW signals are present in the data stream simultaneously, i.e., the overlapping
signals~\cite{Regimbau:2009rk, Samajdar:2021egv, Pizzati:2021apa,
Relton:2021cax, Wu:2022pyg, Wang:2025ckw}.

The overlapping of signals is one of the main challenges in the future GW data
analysis~\cite{Abac:2025saz}. When extracting information from one signal, the
existence of other signals may bias the parameter estimation (PE) for the GW
source~\cite{Samajdar:2021egv, Pizzati:2021apa, Himemoto:2021ukb,
Janquart:2022fzz, Wang:2023ldq}, thus affecting further studies, such as
introducing fake evidence for binary precession or general-relativity
violation~\cite{Relton:2021cax, Hu:2022bji, Dang:2023xkj}.  On the other hand, a
global fit, or called the joint PE, for all signals above the detection
threshold is computationally expensive~\cite{Janquart:2022fzz, Baka:2025yqx},
while the signals below the threshold can still lead to 
biases~\cite{Relton:2021cax, Wang:2023ldq}, or form as a confusion
noise~\cite{Regimbau:2009rk, Antonelli:2021vwg, Wu:2022pyg, Reali:2022aps,
Johnson:2024foj}.  In the future space-based GW detectors, such as
LISA~\cite{LISA:2017pwj}, Taiji~\cite{Hu:2017mde} and
TianQin~\cite{TianQin:2015yph, Gong:2021gvw} programs, the issue of overlapping
signals is even more severe due to their lower frequency band, where numerous
signals are present simultaneously~\cite{Ruiter:2007xx, Belczynski:2008nh,
Niu:2024wdi}.

There have been some studies developing methods to identify and analyze the
overlapping signals in the XG detectors.  Past explorations include the
hierarchical subtraction~\cite{Antonelli:2021vwg, Janquart:2022fzz}, Hough
transform~\cite{Miller:2023rnn}, machine-learning based
algorithms~\cite{Langendorff:2022fzq, Alvey:2023naa, Papalini:2025exy}, and 
their combinations~\cite{Hu:2025vlp}. To reduce the computational burden of
joint PE, techniques such as relative binning~\cite{Zackay:2018qdy,
Leslie:2021ssu} are employed~\cite{Baka:2025yqx}. In these methods, a clear
understanding of how the overlap between signals affects PE is crucial, which
can be further characterized by the biases in the single-signal PE
(SPE)~\cite{Samajdar:2021egv, Relton:2021cax, Antonelli:2021vwg, Wang:2023ldq}.
For example, the criteria for overlapping signals are set based on the magnitude
of the biases, facilitating the estimation of overlapping
rates~\cite{Pizzati:2021apa, Relton:2021cax, Wang:2025ckw}.  Also, the SPE
biases naturally arise in the hierarchical-subtraction method where the signals
are extracted and removed one by one according to their signal-to-noise ratios
(SNRs)~\cite{Janquart:2022fzz}.  Furthermore, the biases in the SPE serve as
good indicators for the necessity of a joint PE~\cite{Baka:2025yqx}.  The
correlation coefficients between the parameters of different signals in a joint
PE~\cite{Antonelli:2021vwg, Pizzati:2021apa, Johnson:2024foj} can also be used
to characterize the influence of overlapping signals, which are found to have
similar behaviors as the SPE biases~\cite{Wang:2023ldq}.

Due to the high dimensionality of the parameter space, it is a challenging work
to explore the formation of biases and their dependence on the parameters of 
overlapping signals, and there is a trade-off between the level of detail and
the scope that can be discussed.  Several studies conducted the standard
Bayesian inference and calculated the full posterior distribution for some
representative cases~\cite{Samajdar:2021egv, Relton:2021cax, Pizzati:2021apa},
which faithfully showed the impact of overlapping but were limited to the number
of examples.  Some approximations, such as the Fisher-matrix (FM) method, were
adopted to extend the study to a wider parameter space~\cite{Antonelli:2021vwg,
Pizzati:2021apa, Himemoto:2021ukb, Johnson:2024foj}, allowing for tracing the
biases over some parameters. Another approach is establishing approximate,
(semi-)analytical formulae for the biases and examining the parameter dependence
explicitly~\cite{Himemoto:2021ukb, Wang:2023ldq}, where the behaviors of the
biases can be discussed in a theoretical view with physically-motivated
interpretations. Recently, \citet{Wang:2023ldq} have shown in detail and
quantitatively that the biases originate from the overlap of the frequency
evolution between signals, \Referee{based on which some behaviors of the biases found in
previous studies are well explained, such as the asymmetry distribution with
respect to the merger order and the oscillatory dependence on the
time separation. These behaviors are revisited and further investigated
in Sec.~\ref{subsec:bias in
complex plane} of this work.}

Currently, most studies focus on biases and their dependence on intrinsic
parameters such as the masses and spins of the binary, along with some extrinsic
parameters like luminosity distance and coalescence time and
phase~\cite{Samajdar:2021egv, Pizzati:2021apa, Antonelli:2021vwg, Wang:2023ldq}.
However, the dependence on the remaining extrinsic parameters is not fully
explored, including the sky location $\sky$ (the right ascension $\alpha$ and
declination $\delta$), the polarization angle $\psi$, and the inclination angle
$\iota$. These parameters describe the orbital orientation of the binary, and
become  important in a detector network.  \citet{Pizzati:2021apa} placed GW
sources with the same sky location and orbital orientation, where the biases
behaved similarly to those considered in a single detector with an
angle-averaged response function~\cite{Antonelli:2021vwg, Wang:2023ldq}. 
Indeed, for each individual detector in the network, the overlapping signals are
just like those in a single detector as studied before. \Referee{For example,
one may expect that the different arrival times of the signals in different
detectors, due to their different locations, may help separate the overlapping
signals in time and alleviate the biases. However, the triangulation of sky
location may also be biased by the signal overlap, thereby making the PE worse.}
  \citet{Relton:2021cax} discussed the impact of the
network explicitly, where they showed two representative cases in the LIGO-Virgo
network using full Bayesian inference: (i) the best case scenario where two
signals are at opposite sides of the line of sight between detectors, and (ii)
the worst case where they have the same sky location.  As expected, the biases
in the worst case are more significant since signals have the same arrival time
differences in different detectors and interfere in the same way.  The two
scenarios have the extremes of the time separation between signals in the
network, while there is still a lack of a comprehensive understanding of how the
general sky location and orbital orientation of the overlapping signals affect
the biases in a network.  In addition, considering the overlapping signals
beyond a single detector is of practical relevance, as both current and future
GW observations are operated in a detector network.

In this work, we extend the case study by \citet{Relton:2021cax} to
systematically investigate the behaviors of the SPE biases in a network of XG
detectors. Specifically, we focus on the following questions: 
%--
\begin{enumerate}[(i)]\setlength\itemsep{0em}
    \item For a detector in the network, how do the biases depend on the
    location and orientation of both the detectors and the GW source?
    \item How do these angle parameters affect the combination of biases from
    individual detectors into the network?
    \item Compared to the single-detector case, will the network increase or
    decrease the biases in general? 
\end{enumerate}
%--
To efficiently explore the parameter space, we adopt the FM
method~\cite{Finn:1992wt, Vallisneri:2007ev} to calculate the biases, which has
been adopted in many previous studies~\cite{Pizzati:2021apa, Himemoto:2021ukb,
Hu:2022bji, Dang:2023xkj} and found to agree well with the full Bayesian
inference in the SPE of overlapping signals~\cite{Antonelli:2021vwg,
Wang:2023ldq}. 

Following \citet{Wang:2023ldq}, we analyze the bias behaviors in the network
from a (semi-)analytical perspective.  To illustrate the network's impact more
directly, we consider the same overlap between two non-spinning binary black
hole (BBH) signals discussed in \citet{Wang:2023ldq} in a 3-detector network.
\Referee{We also adopt the same variable set in the SPE, including the chirp mass, symmetric
mass ratio, luminosity distance, and coalescence time, and study how the biases
of these parameters depend on different but fixed geometric angle parameters.}
We introduce a new complex quantity, the \textit{bias integral} $J$, which is
particularly useful for characterizing these biases and their dependence on
extrinsic parameters.  The biases can be expressed as a linear combination of
the real parts of the bias integrals for varying parameters.  For example, we
show both numerically and analytically that the bias integral rotates in the
complex plane with a slowly varying amplitude as the time separation between
signals increases.  This provides a theoretical explanation for the oscillatory
behaviors of the biases observed in previous studies~\cite{Pizzati:2021apa,
Himemoto:2021ukb, Wang:2023ldq}.  The bias integral allows for a more systematic
study of how angle parameters affect biases while offering a clear physical
interpretation. 

We summarize our main conclusions as answers to the above three questions:
%--
\begin{enumerate}[(i)]\setlength\itemsep{0em}
    \item
    We explicitly show that how---through the geometric angle parameters---the
    bias behaviors in the $D$-th detector, $J_D$, is related to a reference bias
    integral $\bar{J}$, which is understood as the bias integral at the Earth
    center with an angle-averaged response function.  The different locations of
    the detectors correspond to the arrival time differences in the bias
    integral, while their different orientations effectively multiply the bias
    integral by a constant complex factor which depends on the angle parameters.
    \item The bias integral in the network, $J^{\rm net}$, is the sum of those
    $J_D$ in individual detectors. Due to the different locations and
    orientations of the detectors, $J_D$ has different amplitudes and phases
    depending on the angle parameters.  Similarly to the wave optics, this leads
    to a constructive or destructive combination of the bias integrals in the
    network. When the time separation between signals increases, the geometric
    effect from the detectors' locations decreases, while the effect from
    different responses remains unchanged. Only for small separations ($\lesssim
    0.2\,{\rm s}$), the location effect dominates the combination of $J^{\rm
    net}$. 
    \item We simulate a population of overlapping signals with fixed intrinsic
    parameters, while have isotropic sky locations and orientations, and
    uniformly distributed time separations and coalescence phases. We find that
    nearly half ($\gtrsim 40\%$) of the overlapping signals result in larger
    biases in the network compared to that in a single detector. This holds for
    overlapping signals with time separation thresholds ranging from $0.1\,{\rm
    s}$ to $5\,{\rm s}$. This fraction increases with the threshold, as more
    samples with larger time separations mitigate the effect of the detectors'
    locations. 
\end{enumerate}
%--
Overall, our work provides a comprehensive understanding of how the geometric
angle parameters $\{\alpha, \delta, \psi,\iota\}$, as well as the coalescence
time $t_c$ and phase $\phi_c$, of overlapping signals affect the SPE biases in
the detector network, serving as a natural and useful complement to previous
studies that focused on bias behaviors in a single detector. The newly
introduced bias integral $J$ is a powerful tool for characterizing the biases,
which will aid in the development of methods for analyzing overlapping signals.
We also highlight the risk that neglecting other signals can lead to even larger
biases in a detector network than in a single detector.  This may contradict
from the simple expectation that networks always help separate signals and
suppress biases. It is because that the network amplifies all received signals
equivalently, reducing statistical uncertainty and making biases more
significant. 

This paper is organized as follows. In Sec.~\ref{sec:PE methods}, we introduce
the standard Bayesian PE in the GW data analysis and the FM method for
calculating the biases.  In Sec.~\ref{sec:analytical expression}, we introduce
the bias integral, and show its dependence on the extrinsic parameters
explicitly. We also show the relationship between the bias integral in a single
detector and that in the network.  Section~\ref{sec:case studies} presents case
studies to show how the detectors' locations and orientations affect the
combination of biases in the network.  In Sec.~\ref{sec:average over extrinsic
parameters}, we compare the biases in the network and in a single detector at
the population level, averaging over the extrinsic parameters.
Section~\ref{sec:summary and discussion} concludes with a summary and some
discussions. 

%---------------------------------------------------------------------
\section{Parameter Estimation Methods}\label{sec:PE methods}
%---------------------------------------------------------------------

%---------------------------------------------------------------------
\subsection{Bayesian data analysis of GWs}
%---------------------------------------------------------------------

The main task of PE is to extract the parameters of the GW source, $\bmt$, from
the observed data, $g$. In the Bayesian framework, this is done by finding the
posterior distribution of the parameters given the data, $P(\bmt|g)$.  According
to Bayes' theorem, $P(\bmt|g)$ is expressed as 
%----
\begin{equation}
P(\bmt|g) = \frac{P(g|\bmt)P(\bmt)}{P(g)} =
\frac{L(\bmt)\pi(\bmt)}{Z}\,,\label{eq:bayes theorem}
\end{equation}
%----
where in the second equality, we have introduced the terminologies in the
Bayesian inference: $\pi(\bmt)\equiv P(\bmt)$ is the \textit{prior},
representing the knowledge of parameters before observing the data; the
\textit{likelihood} $L(\bmt) \equiv P(g|\bmt)$ is a function of $\bmt$ given
$g$, describing how well the parameters fit the data within the adopted model
(not shown explicitly here); the denominator $Z \equiv P(d)$ is the
\textit{evidence}, which can also be regarded as the normalization factor of the
numerator according to the law of total probability, $Z = \int L(\bmt)\pi(\bmt)
\rmd\bmt$.  \Referee{
In the right-hand side of Eq.~\eqref{eq:bayes theorem}, the prior is given beforehand, and the
    likelihood is determined by the data, thus in principle the posterior $P(\bm \theta|g)$
    can be calculated accordingly.}

In the GW data analysis, the data $g$ are regarded as the sum of the GW signal
$h$ and the detector noise $n$.  Assuming that the noise is stationary and
Gaussian with zero mean, the likelihood in a single detector can be expressed
as~\cite{Finn:1992wt}
%----
\begin{equation}
	L(\bmt) \propto e^{-\frac{1}{2}(g-h,\, g-h)}\,,\label{eq:likelihood in
	single detector}
\end{equation}
%----
with the inner product defined by
%----
\begin{equation}
	(u\,,v):= 4 \Re \int_{0}^{\infty} \frac{u^{*}(f) v(f)}{S_{n}(f)} {\rm d}
	f\,, \label{eq:inner product}
\end{equation}
%----
for any two data streams, $u$ and $v$. In Eq.~(\ref{eq:inner product}), $S_n(f)$
is the one-sided power spectral density (PSD) of the noise, and $u(f)$ and
$v(f)$ are the frequency-domain representations of $u$ and $v$, respectively.

In a network of detectors, assuming that the noise in each detector is
independent, the whole likelihood can be expressed as the product of the
likelihoods in each detector.  Using  boldface for  quantities in the network,
e.g., ${\bm u}\equiv (u_1, u_2, \ldots, u_{N_D})$ for a network with $N_D$
detectors, we define the network inner product as
%----
\begin{equation}
(\bmu\,,\bmv) := \sum_{D} (u_D\,,v_D)_D = \sum_{D} 4\Re
\int\frac{u_D^*(f)v_D(f)}{S^D_n(f)}\rmd f\,,
\end{equation}
where $(\cdot\,,\cdot)_D$ is the inner product in the $D$-th detector, and
$S^D_n(f)$ denotes the corresponding PSD.  Then, the network likelihood is
%----
\begin{equation}
L(\bmt) \propto e^{-\frac{1}{2}(\bmg-\bmh,\, \bmg-\bmh)}\,,
\label{eq:likelihood in network}
\end{equation}
%----
with data $\bmg$ and signal model $\bmh$ in their vector forms. In addition, the
network SNR for signal $\bmh$ is $\rho := \sqrt{(\bmh\,,\bmh)}$.

%---------------------------------------------------------------------
\subsection{Fisher-matrix method for calculating biases}
%---------------------------------------------------------------------

\Referee{Given the observed GW data $\bmg$ and the prior $\pi(\bmt)$, in principle the
posterior $P(\bmt|\bmg)$, as a distribution function of the parameters $\bmt$,
can be computed according to Eqs.~(\ref{eq:bayes
theorem}--\ref{eq:likelihood in network}).
 In practice, this is usually
substituted by drawing samples from the posterior using stochastic sampling
algorithms,} such as the Markov-chain Monte Carlo
method~\cite{Christensen:1998gf, Christensen:2004jm, Sharma:2017wfu} and nested
sampling~\cite{Skilling:2004pqw, Skilling:2006gxv}. However, these methods are
still computationally expensive, especially when lots of inferences are
required.  In this work, we seek for a comprehensive and (semi-)analytical
understanding of how the detector network affects the PE of overlapping GWs,
instead of investigating the posterior distribution in detail for some specific
cases.  Therefore, we follow~\citet{Wang:2023ldq} and adopt the FM method to
approximate the posterior, and focus on the biases in the SPE due to the
existence of other signals.  Similar treatments have also been applied to deal
with lots of PEs in literature~\cite{Antonelli:2021vwg, Himemoto:2021ukb,
Pizzati:2021apa, Hu:2022bji}. 

Specifically speaking, the FM in the GW data analysis is defined
as~\cite{Finn:1992wt}
%----
\begin{equation}
	F_{\alpha\beta} :=
	\left(\bmh_{,\alpha}\,,\bmh_{,\beta}\right)\,,\label{eq5:FM}
\end{equation}
%----
where $\bmh_{,\alpha}$ is the partial derivative of the signal model with
respect to the $\alpha$-th parameter and evaluated at injected parameters.  If
the data $\bmg$ are indeed generated by the signal model $\bmh$, in the large-SNR
limit, the posterior tends to a multivariate Gaussian distribution, whose
covariance matrix is given by the inverse of the FM, denoted as $\Sigma =
F^{-1}$.  Then, the standard deviation of the $\alpha$-th parameter is given by
$\Delta \theta^\alpha_{\rm stat} = \sqrt{\Sigma^{\alpha\alpha}}$, representing
the statistical uncertainty of the inference due to the noise. 

Similarly to~\citet{Wang:2023ldq}, we consider the overlapping of two signals,
$\bmh^{(1)}$ and $\bmh^{(2)}$, and investigate how the second signal affects the
SPE of the first signal. The likelihood now only depends on $\bmt^{(1)}$, 
%----
\begin{equation}\label{eq:SPE likelihood} 
	L\big(\bmt^{(1)})\propto \exp\left\{-\frac{1}{2}
	\left(\bmg-\bmh^{(1)}\big(\bmt^{(1)}\big)\,,
	\bmg-\bmh^{(1)}\big(\bmt^{(1)}\big)\right)\right\}\,,
\end{equation}
%----
while the data $\bmg$ contain the contributions from both signals and the noise.
Then, we adopt the PE biases to characterize the influence of the second signal,
$\Delta \theta^\alpha_{\rm bias}:= {\rm{E}}[\hat{\theta}^\alpha]
-\tilde{\theta}^\alpha $, where $\hat{\theta}^\alpha$ and
$\tilde{\theta}^\alpha$ are the inferred and injected values of the $\alpha$-th
parameter, respectively, and ${\rm{E}}[\cdot]$ denotes the average over all
noise realizations.  In the linearized-signal approximation (LSA), which is
equivalent to the FM method in the large-SNR limit~\cite{Vallisneri:2007ev}, the
biases can be analytically expressed as $\Delta \theta^\alpha_{\rm bias} =
\Sigma^{\alpha\beta} (\bmh^{(1)}_{,\beta}\,, \bmh^{(2)})$, where all quantities
are evaluated at the injected parameters, $\tilde{\bmt}^{(1)}$ and
$\tilde{\bmt}^{(2)}$, and $\Sigma$ corresponds to the FM only for the first
signal, i.e., $F_{\alpha\beta} = \big(\bmh^{(1)}_{,\alpha}\,,
\bmh^{(1)}_{,\beta}\big)$~\cite{Wang:2023ldq}. 

Next, we compare the biases introduced by the second signal with statistical
uncertainties due to the noise, and define the (dimensionless) reduced bias as
\begin{equation}
B_\alpha:= \frac{\Delta \theta^\alpha_{\rm bias}}{\Delta \theta^\alpha_{\rm
stat}} = \frac{\Sigma^{\alpha\beta} \big(\bmh^{(1)}_{,\beta}\,,
\bmh^{(2)}\big)}{\sqrt{\Sigma^{\alpha\alpha}}}\,,\label{eq:reduced bias}
\end{equation}
where only $\beta$ is summed. The reduced bias has a clear physical explanation:
$|B_\alpha|\ll 1$ means that the influence of the second signal is hidden in the
random noise and can be safely ignored, while $|B_\alpha|\gg 1$ means that
ignoring the fact of overlapping of  two signals will lead to a statistically
significant bias in  data analysis.  Since all quantities are evaluated at the
injected parameters, in the followings we omit the tilde notation for
simplicity.

The above formulae in this subsection, along with their derivation, are formally
the same as those in Ref.~\cite{Wang:2023ldq}, except that the network inner
product is the sum of the inner products in each detector, as shown in
Eq.~\eqref{eq:inner product}.  Based on this, we can firstly sketch out how the
network affects the (reduced) biases:
%----
\begin{enumerate}[(i)]
	\item The network increases the SNR of signals, $\rho_{\rm net}^2 = \sum_D
	\rho_D^2$ with $\rho_D$ being the SNR in the $D$-th detector. It increases
	the reduced biases~\cite{Antonelli:2021vwg, Wang:2023ldq}.
    \item Due to different locations and orientations of the detectors, the
    inner products in different detectors,
    $\big(h^{(1)}_{D,\alpha}\,,h^{(2)}_D\big)_D$, will be different, and they
    interfere constructively or destructively in calculating $\big(\bm
    h^{(1)}_{,\alpha}\,,\bmh^{(2)}\big)$.
\end{enumerate}
%----
In the next section, we explicitly show the dependence of the inner products on
the parameters of the overlapping signals, and demonstrate the role of these
parameters in summing up $\big(h^{(1)}_{D,\alpha}\,,h^{(2)}_D\big)_D$ from an
analytical view.

%---------------------------------------------------------------------
\section{Analytical Expression for Biases in the Network}
\label{sec:analytical expression}
%---------------------------------------------------------------------

%---------------------------------------------------------------------
\subsection{GW signals and parameters in the network}
%---------------------------------------------------------------------

\citet{Wang:2023ldq} considered the overlapping of two non-spinning BBH signals,
and adopted the \textsc{IMRPhenomD} waveform template~\cite{Husa:2015iqa,
Khan:2015jqa}.  In that work, the variable parameters are the chirp mass $\cal
M$, symmetric mass ratio $\eta$, luminosity distance $d_L$, and the coalescence
time $t_c$; the waveform is averaged over the sky location $\sky$ (or
equivalently the right ascension $\alpha$ and declination $\delta$), the
inclination angle $\iota$, and the polarization angle $\psi$, corresponding to
GWs recorded by a detector at the Earth center with an angle-averaged antenna
response function.  Here we further investigate how these geometric angle
parameters, along with the merger time $t_c$ and phase $\phi_c$ in the
Earth-center frame, affect the PE biases in a detector network.  We divide the
signal parameters $\bmt$ into two parts, i.e., $\bmt =
\big\{\sky,\psi,\iota,\phi_c,t_c, \bmxi\big\}$: the first part includes
$\big\{\sky,\psi,\iota,\phi_c,t_c\big\}$, while the second part includes the
remaining parameters, such as $\cal M$, $\eta$, and $d_L$, denoted as $\bmxi$
collectively.

We
explicitly show the dependence of the signal $h_D$ on these geometric
parameters~\cite{Veitch:2009hd, Maggiore:2007ulw},
%----
\begin{widetext}
\begin{align}
h_D(f;\bmt) = & \Big[F^D_+(\sky,\psi)h_+(f;\bmxi)\frac{1+\cos^2 \iota}{2} +
F^D_\times(\sky,\psi)h_\times(f;\bmxi)\cos \iota \Big] e^{i\phi_c-2\pi i f
\big[t_c+\tau_D(\sky)\big]}\,, \label{eq:signal in detector} 
\end{align}
\end{widetext}
%----
where $F^D_+$ and $F^D_\times$ are the antenna pattern functions of the $D$-th
detector whose expressions can be found in~\citet{Maggiore:2007ulw}, and
$\tau_D$ is the time delay from the Earth center to the $D$-th detector,
$\tau_D(\sky) = -{\bm r}_D \cdot \sky/c$, where $c$ is the speed of light. In
Eq.~(\ref{eq:signal in detector}), $h_+$ and $h_\times$ are the plus and cross
polarizations of the incoming GW, which do not depend on the detector index $D$.
\Referee{In the above expression, we ignore the effects due to Earth rotation and the
finite-size of detectors, thus the pattern function and time delay are treated as
constants, which is valid for short signals~\cite{Essick:2017wyl, Chen:2024kdc}.} To further highlight
the contributions of the geometric parameters to the GW signal, we express $h_+$
in terms of its amplitude and phase, $h_+ = \mathcal{A} e^{i\phi}$. Noting that
$h_\times = -i h_+$. Equation~\eqref{eq:signal in detector} can be rearranged
as,
%----  
\begin{equation}
    h_D(f;\bmt) =a_D {\cala}e^{i\phi+i(\phi_c+\varphi_D)-2\pi i f
    (t_c+\tau_D)}\,,   \label{eq:signal in detector rearranged}
\end{equation}
%----
where $a_D$ and $\varphi_D$ are functions of $\sky$, $\psi$, and $\iota$,
satisfying $ a_De^{i\varphi_D} = F^D_+(1+\cos^2 \iota)/2-iF^D_\times\cos \iota$.
From the above equation, one clearly see that the geometry angles affect the
signal via three quantities: $a_D$, $\varphi_D$, and $\tau_D$, and the latter
two further contribute to the signal after adding with $\phi_c$ and $t_c$,
respectively.

%---------------------------------------------------------------------
\subsection{Bias integral and its dependence on geometric parameters}
\label{subsec:bias integral}
%---------------------------------------------------------------------

Based on the signal model in Eq.~\eqref{eq:signal in detector} or
\eqref{eq:signal in detector rearranged}, one can obtain the reduced biases for
given overlapping signals with parameters $\bmt^{(1)}$ and $\bmt^{(2)}$ using
Eq.~\eqref{eq:reduced bias}.  Before diving into the numerical calculations, we
first try to analytically show how the geometric parameters contribute to the
biases.  As in \citet{Wang:2023ldq}, the second signal contributes to the
reduced bias, $B_\alpha$, only via the inner product
$\big(\bmh^{(1)}_{,\alpha}\,,\bmh^{(2)}\big)$, which now is the summation of the
individual inner products in each detector. When dealing with the summation, we
find that it is quite useful to define the so-called \textit{bias integral},
%----   
\begin{equation}
    J_{D\alpha} := 4\int \frac{\big(A^{(1)}_{D,\alpha}
    -i\Phi^{(1)}_{D,\alpha}A^{(1)}_D\big) A^{(2)}_D}{S^D_n}
    e^{i(\Phi^{(2)}_D-\Phi^{(1)}_D)}\rmd f\,,\label{eq:bias integral}
\end{equation}
%----
where $A_D$ and $\Phi_D$ are the amplitude and phase of the signal in the $D$-th
detector, i.e., $h_D = A_D e^{i\Phi_D}$.  Mathematically, the bias integral is
just the complex quantity in the inner product before taking the real part. 
However, one will find later that it allows us to conveniently relate
$J_{D\alpha}$ in different detectors, and separate out the contributions of the
geometric parameters.  According to the waveform in Eq.~\eqref{eq:signal in
detector rearranged}, we have $A_D = a_D \cala$ and $\Phi_D = \phi + \phi_c+\varphi_D -
2\pi f(t_c+\tau_D)$. 

Here we are more interested in comparing the bias behaviors in a single detector
with that in a network.  Therefore, we choose the same variable parameters as
in~\citet{Wang:2023ldq}, $\big\{{\cal M},\eta,d_L,t_c\big\}$, and discuss how
the extrinsic geometric parameters $\big\{\sky,\psi,\iota,t_c,\phi_c\big\}$
affect their biases. This brings another benefit, that the derivatives of $a_D$,
$\varphi_D$, and $\tau_D$ with respect to these parameters are zero.  \Referee{For $\cal
M$, $\eta$, and $d_L$, they can be included in $\bmxi$ collectively, and the
derivatives with respect to these parameters are $A_{D,\alpha} =
a_D\cala_{,\alpha}$ and $\Phi_{D,\alpha} = \phi_{,\alpha}$. For the coalescence
time $t_c$, we have $A_{D,\alpha} = 0$ and $\Phi_{D,\alpha} = -2\pi f$.}
Substituting them into the bias integral, we rewrite $J_{D\alpha}$ as
%----
\begin{equation}
    J_{D\alpha} = \frac{a^{(1)}_Da^{(2)}_D}{\bar{a}^2} \bar{J}_\alpha(\Delta t_c
    +\Delta \tau_D)e^{i\Delta \varphi_D}\,,\label{eq:bias integral from J_bar}
\end{equation}
%----
where $\Delta$ denotes the difference between the two signals, e.g., $\Delta t_c
\equiv t_c^{(2)}-t_c^{(1)}$, and $\bar{a}^2$ is the averaged value of $a_D^2$
over the sky location $\sky$, the polarization angle $\psi$, and the inclination
angle $\iota$; $\bar{a}^2 = 4/25$ for L-shape interferometer
detectors~\cite{Maggiore:2007ulw, Liu:2020nwz}. For other designs of detectors,
such as the triangle configuration, the averaged value of $a_D^2$ can be
different, but Eq.~\eqref{eq:bias integral from J_bar} can still be used
regarding $\bar{a}^2= 4/25$  as just a constant factor. We also define the
averaged bias integral $\bar{J}_\alpha$ as  
%----
\begin{widetext}
\begin{align}
    \bar{J}_\alpha(\Delta t_c):= 4{\bar{a}}^2\int &
    \frac{\big(\cala^{(1)}_{,\alpha} -i\phi^{(1)}_{,\alpha}\cala^{(1)}
    \big)\cala^{(2)}}{S_n} e^{{i(\Delta\phi+\Delta\phi_c) - 2\pi i f \Delta
    t_c}}\rmd f\,,\label{eq:J_bar}
\end{align}
\end{widetext}
%----
where $\alpha$ runs over parameters in $\big\{{\cal M},\eta,d_L,t_c\big\}$. For
the case of $t_c$, we specifically denote $\cala^{(1)}_{,\alpha}=0$ and
$\phi^{(1)}_{,\alpha} = -2\pi f$ for a consistent representation with  the other
three parameters. Though $\bar{J}_\alpha$ depends on $\bmt^{(1)}$ and
$\bmt^{(2)}$, we specifically mark its dependence on the coalescence time
difference $\Delta t_c$, which helps to distinguish the variable in
Eq.~\eqref{eq:bias integral from J_bar} from the one in Eq.~\eqref{eq:reduced
bias}. \Referee{In addition, here we assume that all detectors have the same noise PSD,
$S_n(f)$, which allows us to relate $J_{D\alpha}$ to a detector-independent
integral $\bar{J}_\alpha$. If a
detector-dependent PSD is chosen, the geometric parameters still only contribute
to $J_{D\alpha}$ via  $a_D$, $\varphi_D$, and $\tau_D$ quantities in
Eq.~\eqref{eq:bias integral from J_bar} while do not affect
Eq.~\eqref{eq:J_bar}, and the descriptions about how these parameters affect the
combination of $J_{D\alpha}$ in different detectors still hold.}  Furthermore, in
Sec.~\ref{subsec:bias in complex plane} one can find that choosing a
detector-dependent PSD mainly changes the modulus of $\bar{J}_\alpha$, which is
less important than its phase in adding up $J_{D\alpha}$ from different
detectors (see below). Therefore, in the following we adopt the same PSD for all
detectors for a simple illustration.

\begin{figure}[t]
\centering
\includegraphics[width=0.35\textwidth]{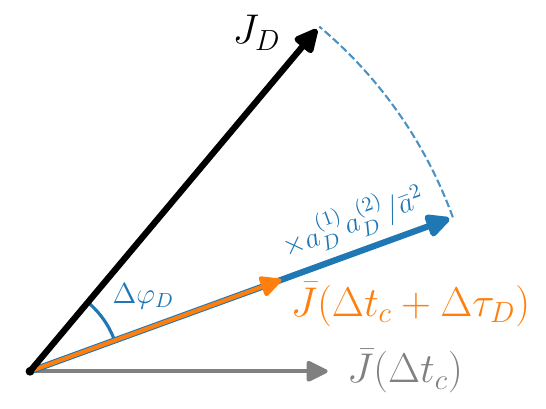}
\caption{Generation of the bias integral $J_{D}$ in the $D$-th detector from the
averaged bias integral $\bar{J}$, showing the roles of the geometric parameters
$\big\{\sky,\psi,\iota\big\}$ according to Eq.~\eqref{eq:bias integral from
J_bar}.  The gray arrow represents $\bar{J}$ as a function of the coalescence
time difference $\Delta t_c = t_c^{(2)}-t_c^{(1)} $ between  two signals. The
orange arrow considers the time delay $\tau_D = -{\bm r}_D \cdot \sky/c$ of the
$D$-th detector from the Earth center, which shifts the variable from $\Delta
t_c$ to $\Delta t_c + \Delta \tau_D$. Blue color shows quantities coming from
the detector response, where $a_D$ and $\varphi_D$ satisfy $ a_De^{i\varphi_D} =
F^D_+(1+\cos^2 \iota)/2-iF^D_\times\cos \iota$ with $F^D_+$ and $F^D_\times$
being the pattern functions of the $D$-th detector which are functions of
$\sky$, $\psi$, and $\iota$. This effect multiplies the orange arrow
$\bar{J}(\Delta t_c + \Delta \tau_D)$ by an amplitude factor
$a_D^{(1)}a_D^{(2)}/\bar{a}^2$ and rotates it by $\Delta \varphi_D$, leading to
the ultimate bias integral in the $D$-th detector, marked in
black.}\label{fig:illustration of bias integral}
\end{figure}

With the bias integral, now we rewrite the inner product in
Eq.~\eqref{eq:reduced bias} as
%----
\begin{equation}
    \big(\bmh^{(1)}_{,\alpha}\,,\bmh^{(2)}\big) = \Re \sum_D
    J_{D\alpha}\,,\label{eq:inner product in terms of bias integral}
\end{equation}
%----
where we have changed the order of the summation and taking the real part in the
definition of the inner product in Eq.~\eqref{eq:inner product}.  This provides
another picture to show how  individual inner products in different detectors
contribute to the total inner product: all $J_{D\alpha}$ can be regarded as
arrows in a complex plane, and the total inner product is the projection of the
sum of these arrows onto the real axis.  In this picture, the role of the
geometric angles, $\big\{\sky,\psi,\iota\big\}$, becomes clear: they only
control the way  to generate $J_{D\alpha}$ in each detector from a same
$\bar{J}_\alpha$ via Eq.~\eqref{eq:bias integral from J_bar}. Similarly to their
effects on the signal in Eq.~\eqref{eq:signal in detector rearranged}, they
contribute to $J_{D\alpha}$ in quantities $a_D^{(1)}$, $a_D^{(2)}$, $\Delta
\varphi_D = \varphi_D^{(2)}-\varphi_D^{(1)}$, and $\Delta \tau_D =
\tau_D^{(2)}-\tau_D^{(1)}$. In Fig.~\ref{fig:illustration of bias integral}, we
illustrate this generation process of $J_{D\alpha}$ in the complex plane and
divide it into two steps. We temporarily ignore the lower script $\alpha$ since
this process is the same for all parameters.  In the first step, one starts with
$\bar{J}(\Delta t_c)$, and finds that considering the time delay of each
detector from the Earth center changes the variable of $\bar{J}$ from $\Delta
t_c$ to $\Delta t_c + \Delta \tau_D$. In the second step, one considers the
location and orientation of a detector, and multiplies $\bar{J}(\Delta t_c +
\Delta \tau_D)$ by an amplitude factor $a_D^{(1)}a_D^{(2)}/\bar{a}^2$ and a
phase shift $e^{i\Delta \varphi_D}$ to obtain $J_{D}$.  The magnitudes of the
time shift, the amplitude factor and the phase shift are numerically discussed
in Sec.~\ref{sec:case studies}.

%----
%----

%---------------------------------------------------------------------
\subsection{Revisiting biases in the complex plane}
\label{subsec:bias in complex plane}
%---------------------------------------------------------------------

Since all bias integral $J_{D\alpha}$ are generated from the same
$\bar{J}_\alpha$, now we discuss the behaviors of $\bar{J}_\alpha$. Though
introduced as an auxiliary function in Eq.~\eqref{eq:bias integral from J_bar},
the physical meaning of $\bar{J}_\alpha$ is clear: it is the bias integral in a
single detector located at the Earth center with the angle-averaged antenna
response. \citet{Wang:2023ldq} adopted the angle-averaged waveform and ignored
the time delay between the detector and the Earth center, which corresponds
exactly to the $\bar{J}_\alpha$ defined here. However, \citet{Wang:2023ldq}
directly studied the inner product $\big(h^{(1)}_{,\alpha}\,,h^{(2)}\big)$,
i.e.\ the real part of $\bar{J}_\alpha$, and found some modulated trigonometric
functions oscillating with $\Delta t_c$.  \Referee{In addition, the analysis used the
AdvLIGO noise PSD, which does not affect the qualitative behavior of the biases
but might impact the typical magnitudes of relevant quantities. For example, since the biases mainly accumulate in the frequency band
where the two signals have time-frequency crossing, a wider sensitive
band in the low-frequency regime will allow significant biases to occur for larger time separations. 
As an extended investigation of~\citet{Wang:2023ldq},
in this subsection we revisit the dependence of $J_\alpha$ on
$\Delta t_c$ in
the complex plane with noise PSDs of the XG detectors.}

Firstly, we note that $\Delta \phi_c$ can be taken out of the integral, that is,
$J_{\alpha}(\Delta \phi_c) = e^{i\Delta \phi_c} J_{\alpha}({\Delta \phi_c = 0})$
for both $J_{D\alpha}$ and $\bar{J}_\alpha$.  Therefore, the effect of the
merger phase is simply to rotate $J_{\alpha}$ anticlockwisely by an angle $\Delta
\phi_c$ in the complex plane. However, this makes the real part of $J_{\alpha}$
vary in $\big[-|J_{\alpha}|,|J_{\alpha}|\big]$ for different $\Delta \phi_c$.
Similarly, we rewrite the reduced bias with the bias integral as
%----
\begin{equation}
    B_{\alpha} = \Re \frac{\Sigma^{\alpha\beta} J_{\beta}}{ \sqrt{\Sigma^{\alpha
    \alpha}}}\,, \quad \max\limits_{\Delta \phi_c} \left| B_\alpha \right| =
    \frac{\big|\Sigma^{\alpha\beta} J_{\beta}\big|}{\sqrt{\Sigma^{\alpha
    \alpha}}}\,,\label{eq:reduced bias in complex plane}
\end{equation}
%----
where we have changed the order of the summation over $\beta$ and have taken the
real part. Since the covariance matrix $\Sigma$ is independent of $\Delta
\phi_c$, the sum $\Sigma^{\alpha\beta}J_{\beta}$ also rotates anticlockwise by
$\Delta \phi_c$ in the complex plane while keeping its modulus unchanged. In
particular, this phenomenon becomes significant for large $|B_\alpha|$, where
the biases are substantial in one injection but nearly zero in another that only
varies $\Delta \phi_c$. Previous studies~\cite{Relton:2021cax, Pizzati:2021apa}
have noted this but did not provide a detailed explanation.  Based on the above
analysis, maybe the modulus of the complex quantity serves as a better
characterization of the biases, representing the ``intrinsic'' influence from
the second signal. This modulus also corresponds to the maximum of $\left| B_\alpha \right|$
over $\Delta \phi_c$, as shown in the second equation of Eq.~\eqref{eq:reduced
bias in complex plane}. In the remaining part of Sec.~\ref{sec:analytical
expression} and Sec.~\ref{sec:case studies}, we fix $\Delta \phi_c = 0$. We will
return to the two characterizations of  biases in Sec.~\ref{sec:average over
extrinsic parameters}, where we simulate overlapping signals with various
extrinsic parameters, including $\Delta \phi_c$.

Now we focus on the dependence of $\bar{J}_\alpha$ on the coalescence time
difference $\Delta t_c$, the most important parameter in analyzing overlapping
signals. \citet{Wang:2023ldq} have shown that the real part of the integrand in
Eq.~\eqref{eq:J_bar} is a modulated trigonometric function oscillating with $f$,
and the integral is dominated by the stable stage where the phase term in the
exponential varies slowly with $f$.  The work also showed that this stable stage
corresponds to a similar stage in the frequency evolution of the two signals, 
which was analytically confirmed by \citet{Johnson:2024foj} with a linear
frequency evolution model.  Furthermore, \citet{Johnson:2024foj} calculated the
Fisher cross-terms between the two signals in the stationary phase approximation
(SPA) limit, and found that the stationary frequency point was very close to the
crossing frequency of the time-frequency tracks of the two signals. Motivated by
these results, here we  adopt SPA to calculate the integral in $\bar{J}_\alpha$.
Namely, we have
%----
\begin{equation}
    \bar{J}_{\alpha} \approx  \bar{J}^{\rm spa}_{\alpha} = {\cal
    Z}_{\alpha}(f^{\rm spa})e^{i\big[\Delta \phi(f^{\rm spa}) - 2\pi f^{\rm spa}
    \Delta t_c\big]} , \label{eq:J_bar with SPA}
\end{equation}
%----
where $f^{\rm spa}$ is the stationary frequency satisfying $\Delta \phi'(f_{\rm
spa}) = -2\pi \Delta t_c$, with prime denoting the derivative with respect to
$f$. $\calz_{\alpha}(f^{\rm spa})$ is defined as  
%----
\begin{align}
    {\cal Z}_{\alpha}(f^{\rm spa}) := & \frac{4\bar{a}^2\big({\cal
    A}^{(1)}_{,\alpha}-i\phi^{(1)}_{,\alpha}{\cal A}^{(1)}\big){\cal
    A}^{(2)}}{S_n}  \sqrt{\frac{2\pi}{|\Delta\phi''|}}e^{\frac{i\pi }{4}{\rm
    sign}(\Delta \phi'')}\,,\label{eq:J_bar amplitude}
\end{align}
%----
where all quantities are evaluated at $f^{\rm spa}$. 

%----
\begin{figure*}[t]
\centering
\includegraphics[width=0.9\textwidth]{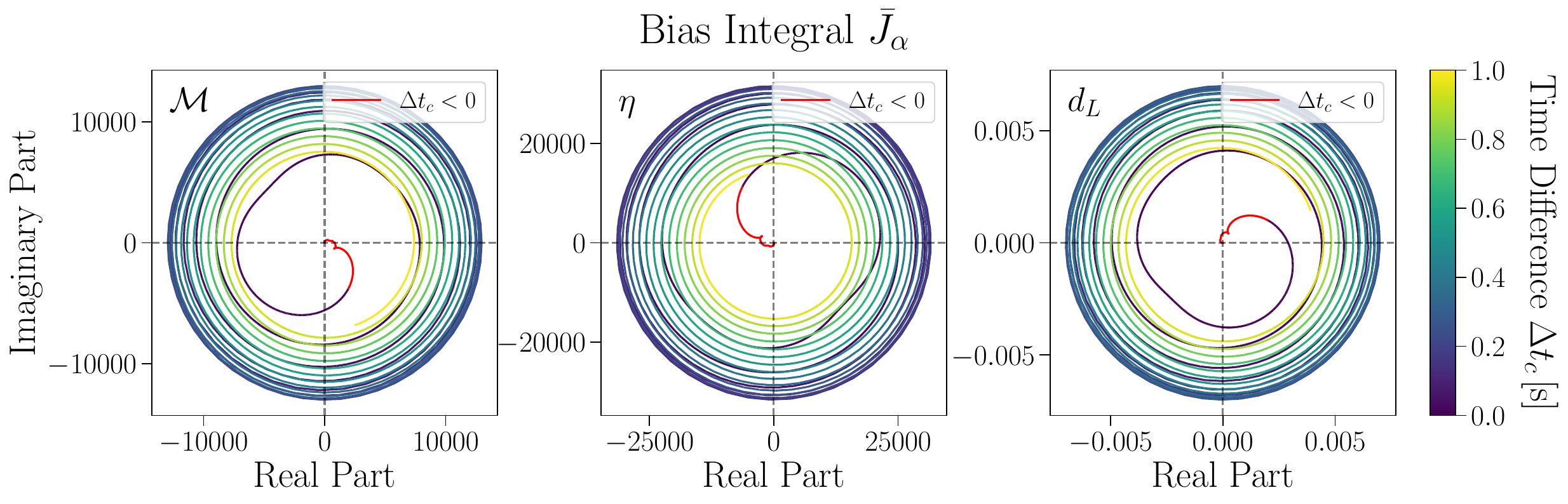}
\caption{Evolution of the bias integral $\bar{J}_\alpha$ in the complex plane,
as a function of the coalescence time difference $\Delta t_c$ for  three
parameters $\cal M$, $\eta$, and $d_L$. The red line marks $\bar{J}_\alpha$
where $\Delta t_c\in[-0.1\,{\rm s},0\,{\rm s}]$, while the remaining part is for
$\Delta t_c\in[0\,{\rm s},1\,{\rm s}]$ and plotted with color bar. Though
$\bar{J}_\alpha$ is dimensional (the reciprocal of $\theta^\alpha$) and has
different scales for different parameters, here we are concerned only about
 its
dependence on $\Delta t_c$. Note that here and below in similar figures, we 
report the bias integral in arbitrary unit because only their relative magnitudes 
are important in the context. We also create an animated version of this figure,
which shows the evolution of $\bar{J}_\alpha$ with $\Delta t_c$ more clearly~\cite{supplemental_material}.}
\label{fig:bias integral}
\end{figure*}
%----

To show how $\bar{J}_\alpha$ varies with $\Delta t_c$, we calculate the
derivative of $\bar{J}^{\rm spa}_\alpha$ with respect to $\Delta t_c$,
%----
\begin{equation}
    \frac{\mathrm{d} \bar{J}^{\mathrm{spa}}_\alpha}{\mathrm{d} \Delta t_c}
    =\left[-2\pi i f - \frac{2\pi}{\Delta \phi''} \frac{1}{\mathcal{Z}_\alpha}
    \frac{\mathrm{d} \mathcal{Z}_\alpha}{\mathrm{d} f}\right]_{f=f^{\rm spa}}
    \bar{J}^{\mathrm{spa}}_\alpha\,.\label{eq:derivative of J_bar}
\end{equation}
%----
In the square bracket, the first term indicates a clockwise rotation of
$\bar{J}_\alpha$ as $\Delta t_c$ increases. Next, we find that $\calz_\alpha$ is
nearly either purely real or imaginary for a $\theta^\alpha$ from $\big\{{\cal
M},\eta,d_L,t_c\big\}$: since $d_L$ only affects the signal amplitude,
$\calz_\alpha$ is real; as $t_c$ only influences the signal phase,
$\calz_\alpha$ is imaginary; for the mass parameters $\cal M$ and $\eta$,
numerical results show that $|\phi_{,\alpha}\cala|\gg
|\cala_{,\alpha}|$~\cite{Wang:2023ldq}, making $\calz_\alpha$ almost purely
imaginary. For these parameters, the second term in the square bracket is nearly
real, representing the change in the amplitude of $\bar{J}_\alpha$. Therefore,
we conclude that the bias integral $\bar{J}_\alpha$ rotates clockwise with
frequency $f^{\rm spa}$, and the modulus of $\bar{J}_\alpha$ is controlled by
$\calz_\alpha$. 

After approximately integrating $\bar{J}_\alpha$ in Eq.~\eqref{eq:J_bar with
SPA}, we take a specific example and numerically discuss $\bar{J}_\alpha$ as a
function of $\Delta t_c$. In Fig.~\ref{fig:bias integral}, we show 
$\bar{J}_\alpha$ for  three parameters, $\cal M$, $\eta$, and $d_L$, with
$\Delta t_c \in [-0.1\,{\rm s}, 1\,{\rm s}]$. In the detector frame, the
parameters of signal 1 are $m^{(1)}_1 = 30\,\mathrm{M_\odot}$, $m^{(1)}_2 =
20\,\mathrm{M_\odot}$, $D_L^{(1)} = 10\,\mathrm{Gpc}$, and the parameters of
signal 2 are $m^{(2)}_1 = 27\,\mathrm{M_\odot}$, $m^{(2)}_2 =
18\,\mathrm{M_\odot}$, $D_L^{(2)} = 60\,\mathrm{Gpc}$. The mass parameters are
the same as in the \textsc{Asymmetric} configuration in~\citet{Wang:2023ldq},
while the distance of the two signals are increased in the light of the XG
detectors.  \Referee{We still
use the \textsc{IMRPhenomD} waveform~\cite{Husa:2015iqa, Khan:2015jqa} to align
with~\citet{Wang:2023ldq}. Adopting the new overhaul model
\textsc{IMRPhenomXAS}~\cite{Pratten:2020fqn} or including
higher-order modes~\cite{Garcia-Quiros:2020qpx, Pratten:2020ceb}
does not affect the overall behaviors of the bias integral significantly. We set the spin parameters of both
signals to zero and
leave a more detailed study including spin effects to future work.}
The parameters above are fixed in examples throughout this work.
The noise PSD is taken as the
CE-2 sensitivity curve~\cite{Reitze:2019dyk, Reitze:2019iox}. In this
configuration, the second signal has a lighter mass, thus the stationary
frequency can exist only for positive $\Delta t_c$, i.e., the second CBC merges
after the first one. If there is no stationary point, Eq.~\eqref{eq:J_bar with
SPA} does not hold, and the exponential term in the bias integral evolves
rapidly with $f$, making the final integral close to zero. In Fig.~\ref{fig:bias
integral}, we mark $\bar{J}_\alpha$ for $\Delta t_c\in[-0.1,0] \,{\rm s}$ in
red, and find that the modulus of $\bar{J}_\alpha$ in this region is indeed
smaller than that in $\Delta t_c\in[0,1] \,{\rm s}$.  For $\Delta t_c\in[0,1]
\,{\rm s}$, the three $\bar{J}_\alpha$ rotate clockwisely when $\Delta t_c$
increases, as we have found in Eq.~\eqref{eq:derivative of J_bar}. The speed of
rotation is much faster than the change in the modulus for most positive $\Delta
t_c$, which indicates that in Eq.~\eqref{eq:derivative of J_bar} the first term
in the square bracket is much larger than the second one. It is also worth
noting that the modulus of $\bar{J}_\alpha$ first increases with $\Delta t_c$
and then decreases. The crossing frequency of the two signals decreases when the
time separation between the two signals increases, so $f^{\rm spa}$ is a
monotonic decreasing function of $\Delta t_c$. Considering that $\calz_\alpha$
is inversely proportional to the PSD $S_n$ at $f^{\rm spa}$, the modulus of
$\bar{J}_\alpha$ will first increase and then decrease as $\Delta t_c$
increases. For  CE-2 PSD, the detector is most sensitive at ${\cal O}(10)\,{\rm
Hz}$, and the maximum of $\bar{J}_\alpha$'s amplitude is reached around $f^{\rm
spa} \approx 20\,{\rm Hz}$ consistently. 

From Eq.~\eqref{eq:derivative of J_bar}, we find that the rotation frequency of
$\bar{J}_\alpha$ is approximately $f^{\rm spa}$, which is the same for all
parameters. Therefore, as $\Delta t_c$ increases, the phase differences between
$\bar{J}_\alpha$ remain approximately constant, which has been numerically
confirmed in Fig.~\ref{fig:bias integral}. The phase difference is determined by
the parameters entering the amplitude or phase of the signal.  Both $\cal M$ and
$\eta$ have a nearly imaginary $\calz_\alpha$, meaning that their
$\bar{J}_\alpha$ will be in phase or differ by $\pi$ (depending on the sign of
$\phi_{,\alpha}$), while $d_L$ has a real $\calz_\alpha$, thus its
$\bar{J}_\alpha$ will have a $\pi/2$ phase difference from that of $\cal M$ and
$\eta$. For $t_c$, it only contributes to the phase of the signal and thus
behaves similarly to $\cal M$ and $\eta$. After considering the sign of
$\phi_{,\alpha}$, $\bar{J}_\alpha$ for $t_c$ has the same phase as
$\bar{J}_\alpha$ for $\cal M$, which we choose not to include  in
Fig.~\ref{fig:bias integral}.

Once taking the real part of $\bar{J}_\alpha$, we obtain the inner product
$\big(h^{(1)}_{,\alpha}\,,h^{(2)}\big)$. Now it is clear that the inner product
is a modulated trigonometric function oscillating with $\Delta t_c$, and the
phase difference between different parameters is also explained concisely and
intuitively. From above discussions, we   note the importance of the modulus of
$\bar{J}_\alpha$, as well as the complex quantity $\Sigma^{\alpha\beta}
J_\beta/\sqrt{\Sigma^{\alpha\alpha}}$ in Eq.~\eqref{eq:reduced bias in complex
plane}, in characterizing the biases. In a realistic PE, there could be
parameters that contribute to the amplitude, the phase, or both parts of the
signal. The bias integral provides a unified way to characterize how the biases
vary across different parameters. 

As a short summary, the inner product in the network can be regarded as the real
part of the sum of the bias integral $J_{D\alpha}$ in different detectors, which
are generated from the same ``fiducial'' integral $\bar{J}_\alpha$ in
Eq.~\eqref{eq:J_bar}. The relationship between $J_{D\alpha}$ and
$\bar{J}_\alpha$ is controlled by the geometric angle parameters
$\big\{\sky,\psi,\iota\big\}$ through $a_D$, $\varphi_D$, and $\tau_D$, with the
expression in Eq.~\eqref{eq:bias integral from J_bar} and the illustration in
Fig.~\ref{fig:illustration of bias integral}. The bias integral $\bar{J}_\alpha$
is a complex quantity that rotates clockwisely with $\Delta t_c$ with a slowly
varying modulus as shown in Eq.~\eqref{eq:derivative of J_bar} and
Fig.~\ref{fig:bias integral}.

%---------------------------------------------------------------------
\section{Case Studies}\label{sec:case studies}
%---------------------------------------------------------------------

%----
\begin{figure*}[t]
\centering
\includegraphics[width=0.8\textwidth]{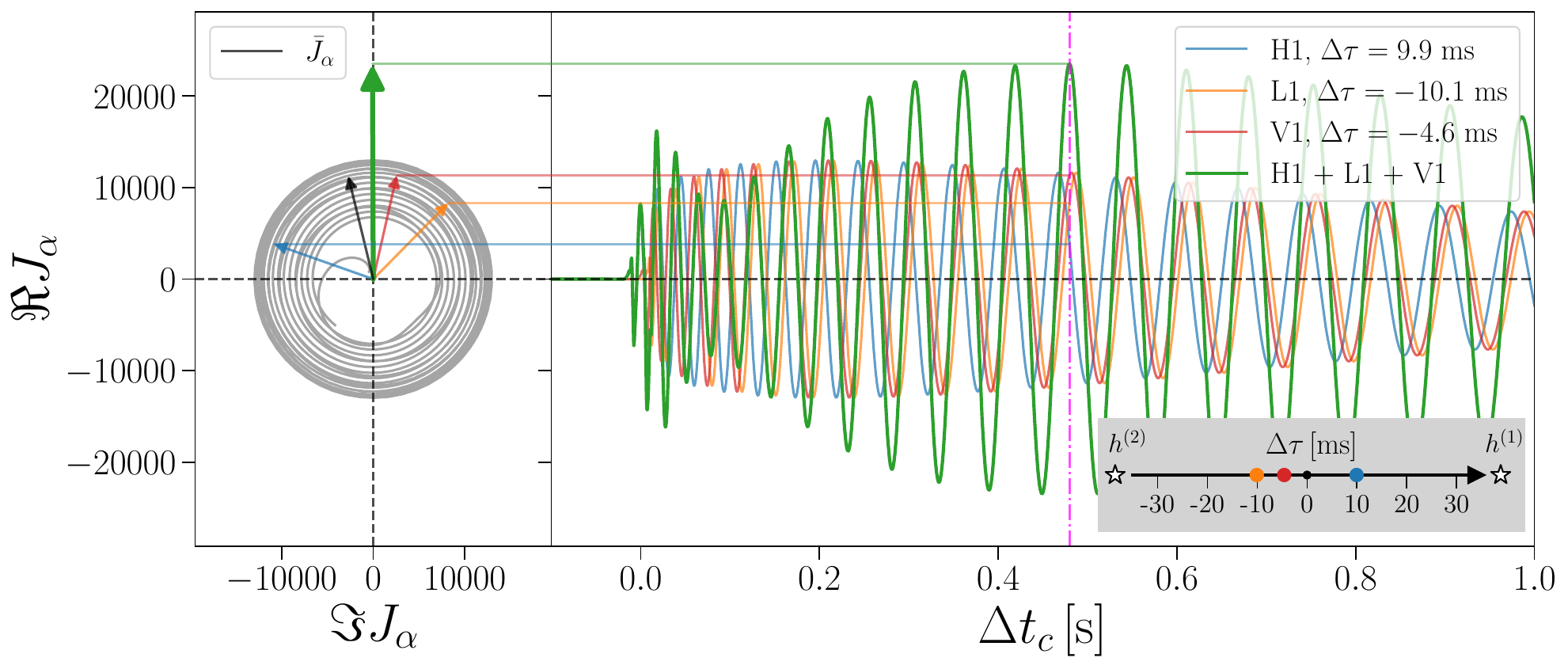}
\caption{Bias integrals in the individual detectors (blue, yellow, and red) and
the whole network (green) for the parameter $\cal M$, only considering the time
delay in different detectors and ignoring their different pattern functions (by
adopting the angle-averaged waveform).  Two sources are located at opposite ends
of the H1$-$L1 line, and the time delays of the three detectors from the Earth's
center are given in the bottom right.  In the left panel, we show the bias
integrals in the complex plane at the maximum point of $\Re J^{\rm net}_\alpha$.
The black arrow represents the averaged bias integral $\bar{J}_\alpha$, while
the gray line shows the evolution track of $\bar{J}_\alpha$ as a function of the
coalescence time difference $\Delta t_c$. The right panel shows the real part of
the bias integral as a function of $\Delta t_c$. The maximum point of $\Re
J^{\rm net}_\alpha$ is marked with a magenta line for reference. 
We also create an animated version of this figure,
which shows the evolution of $\bar{J}_\alpha$ and $J_{D\alpha}$ with $\Delta
t_c$ and their addition more clearly~\cite{supplemental_material}.}
\label{fig:J time delay}
\end{figure*}
%----

In this section, we conduct case studies to show how the network bias integral
$J_{\alpha}^{\rm net}$ forms from the individual bias integrals $J_{D\alpha}$ in
different detectors, and how it depends on the coalescence time difference
$\Delta t_c$. We choose a 3-detector network, located at the LIGO Hanford (H1),
LIGO Livingston (L1), and Virgo (V1) sites, and calculate the time delay and the
antenna pattern functions with the \textsc{pycbc} package~\cite{pycbc}. As
explained in Sec.~\ref{subsec:bias integral}, for simplicity we choose the same
noise PSD for all detectors, which is taken to be the CE-2 sensitivity
curve~\cite{Reitze:2019dyk, Reitze:2019iox} in the frequency range $[3,
3000]\,{\rm Hz}$. The network SNR of  two signals depends on the geometric
parameters $\big\{\sky,\psi,\iota\big\}$. For the angle-averaged waveform over
these parameters, the network SNR of signal 1 and signal 2 are $\rho^{(1)} =
88.7$ and $\rho^{(2)} =13.6$, respectively. As shown in
Fig.~\ref{fig:illustration of bias integral}, we divide the effects of the
geometric parameters into two parts, one is time delay $\tau_D$ coming from
different locations of the detectors, the other is the additional amplitude and
phase factors, $a_D$ and $\varphi_D$, coming from different orientations of the
detectors. In the following, we first discuss the effects of the two parts
separately, and then combine them together.

%---------------------------------------------------------------------
\subsection{Time delay}\label{subsec:time delay}
%---------------------------------------------------------------------

When only considering the time delay $\tau_D$, we adopt the angle-averaged
waveform, and write the bias integral in the $D$-th detector as
$J_{D\alpha}(\Delta t_c) = \bar{J}_\alpha(\Delta t_c + \Delta \tau_D)$ with the
time delay difference $\Delta \tau_D =\tau_D^{(2)}-\tau_D^{(1)}$.  Physically,
$\Delta t_c + \Delta \tau_D$ is the \textit{arrival} time difference of two
signals in the $D$-th detector, which varies across detectors due to their
different geographic locations.

%----
\begin{figure*}[t]
\centering
\includegraphics[width=0.8\textwidth]{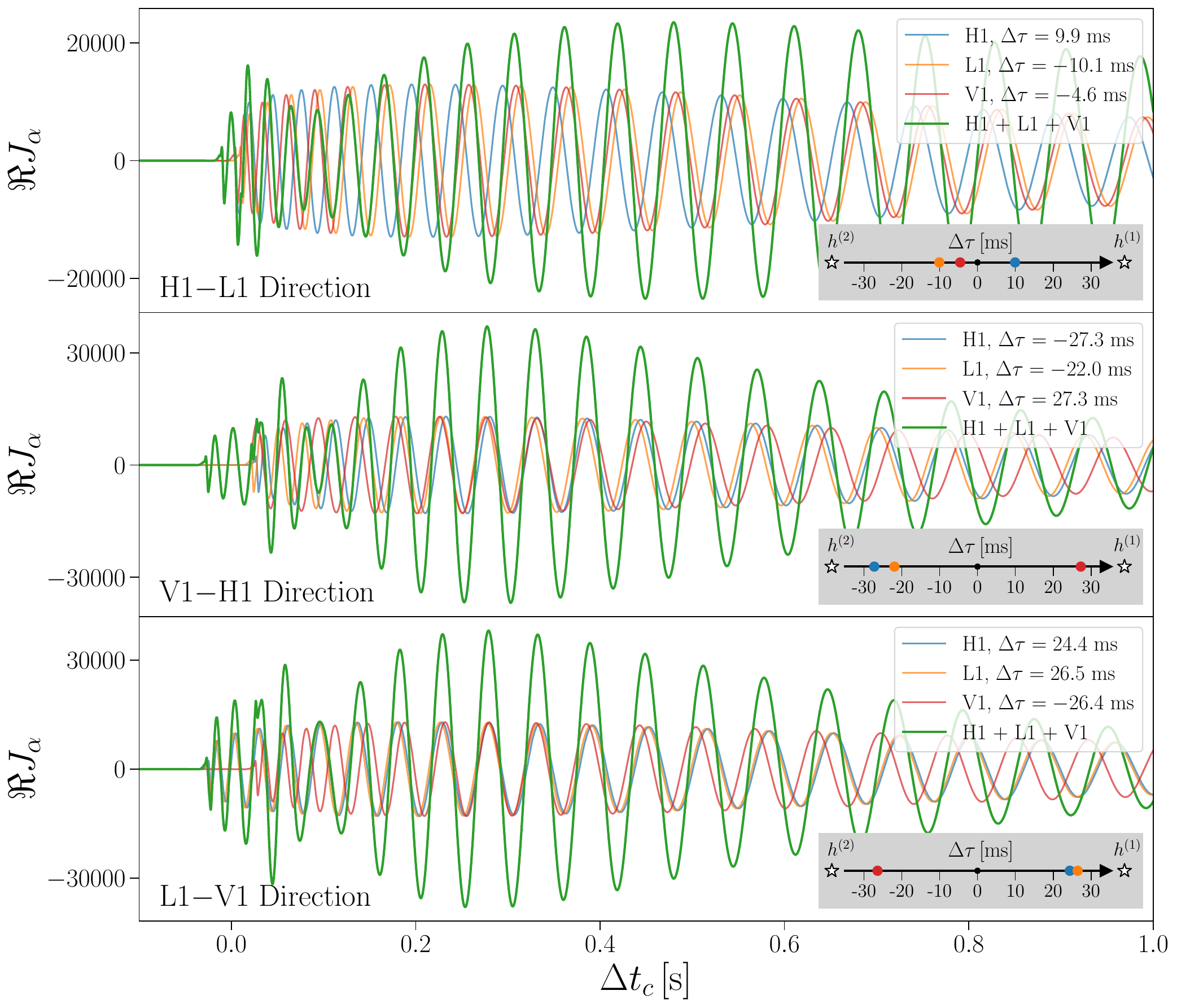}
\caption{The real part of the bias integral as a function of the coalescence
time difference $\Delta t_c$ for the parameter $\cal M$, where only the time
delay effect is considered. From top to bottom, two sources are located at
opposite ends of the H1$-$L1 line (same as Fig.~\ref{fig:J time delay}), V1$-$H1
line, and V1$-$L1 line, respectively.  The colors for individual detectors and
the network are kept the same as in Fig.~\ref{fig:J time delay}.}\label{fig:J
time delay for three cases}
\end{figure*}
%----

In Fig.~\ref{fig:J time delay}, we show the bias integrals $J^{\alpha}$ of the
three detectors and the whole network. We plot their real part, the inner
product $\big(h^{(1)}_{,\alpha}\,,h^{(2)}\big)$, as a function of $\Delta t_c$.
We choose the sky locations of two sources at opposite ends of the H1$-$L1
line. The first source is closer to H1 than L1 when projected onto the H1$-$L1
line, whereas the second source is closer to L1. This arrangement maximizes the
arrival time difference between the two detectors.  We only show the bias
integrals for the parameters $\cal M$, and for remaining three
parameters---$\eta$, $d_L$, and $t_c$---the results are similar. To demonstrate
the relationship between $J_{\alpha}$ and $\big(h^{(1)}_{,\alpha}
\,,h^{(2)}\big)$, we choose $\Re J_{\alpha}$ as the $y$-axis, and $\Im
J_{\alpha}$ as the $x$-axis in the left panel of Fig.~\ref{fig:J time delay},
thus the bias integral anticlockwisely rotates with an increasing $\Delta t_c$
(different from Fig.~\ref{fig:bias integral}).  In this case, $J_{D\alpha}$ of
the three detectors only differ in the variable, rotating along the
track of $\bar{J}_\alpha$ (plotted in gray), and $\Re J_{D\alpha}$ is just the
$\Re \bar{J}_\alpha$ shifted by $\Delta \tau_D$ as shown in the right panel of
Fig.~\ref{fig:J time delay}.  It is important to note that the phase difference
between  three $J_{D\alpha}$ decreases with $\Delta t_c$, since the crossing
frequency decreases for two signals with a larger time separation, and the phase
difference between $J_{D\alpha}$ and $\bar{J}_\alpha$ is approximately $-2\pi
f^{\rm spa} \Delta \tau_D$ in Eq.~\eqref{eq:derivative of J_bar} within the SPA
regime. Therefore, for a large $\Delta t_c$ ($\gtrsim 0.4\,{\rm s}$), the three
$J_{D\alpha}$ arrows tend to align with each other, making $\Re J^{\rm
net}_{\alpha}$ larger than individual $\Re J_{D\alpha}$.  While for a small
$\Delta t_c$ ($\sim 0.1\,{\rm s}$), the three $J_{D\alpha}$ arrows are more
likely to be out of phase and cancel with each other, leading to a smaller $\Re
J^{\rm net}_{\alpha}$.  Physically, for a larger $\Delta t_c$, the additional
time delay $\Delta \tau_D$ becomes less important, and $J_{D\alpha}$ in
different locations tend to be the same $\bar{J}_\alpha$ at the Earth center. 

%----
\begin{figure*}[t]
\centering
\includegraphics[width=0.8\textwidth]{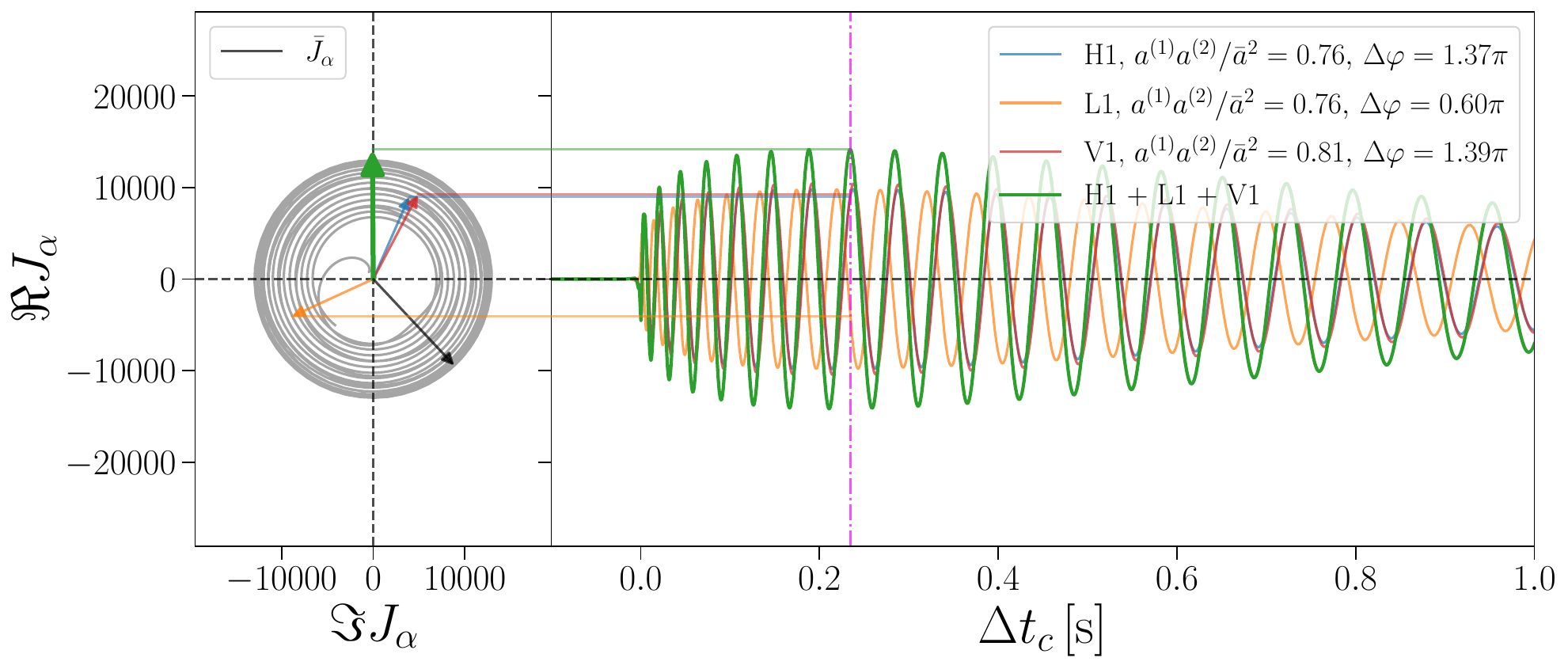}
\caption{Same as Fig.~\ref{fig:J time delay}, but now only considering the
different pattern functions of the detectors while ignoring the time delay
effect. The additional amplitude and phase factors (calculated from pattern
functions) in converting from the averaged bias integral $\bar{J}_\alpha$ to the
detector-dependent $J_{D\alpha}$ are marked. We also create an animated version
of this figure~\cite{supplemental_material}.
}\label{fig:J pattern function}
\end{figure*}
%----

Now we discuss the magnitude of the network bias integral. In the right panel of
Fig.~\ref{fig:J time delay}, we have marked the maximum of $\Re J_{\alpha}^{\rm
net}$ in the right panel (around $0.5\,{\rm s}$), and have plotted the complex
bias integrals in the left panel at the corresponding $\Delta t_c$. This maximum
point can be regarded as the result of two competing effects. On the one hand,
the modulus of $\bar{J}_\alpha$ is decreasing with $\Delta t_c$ ($\gtrsim
0.2\,{\rm s}$), since the crossing frequency of the two signals is decreasing
towards the low frequency cutoff of the PSD.  On the other hand, the phase
difference between the three $J_{D\alpha}$ is also decreasing, which makes them
easier to add up to a larger $\Re J^{\rm net}_{\alpha}$.  The maximum value of
$\Re J_{\alpha}^{\rm net}$ is approximately twice the maximum value of $\Re
\bar{J}_\alpha$. \Referee{To determine the corresponding bias, note that the covariance
matrix $\Sigma$ in the network is exactly one third of that in a single detector
for $N_D = 3$ detectors when adopting the angle-averaged waveform.} To ensure that the reduced bias in the network
exceeds that in a single detector, $|J_{\alpha}^{\rm net}|$ is required to be at
least $\sqrt{3}$ times of $|\bar{J}_\alpha|$, which is satisfied for $\Delta t_c
\gtrsim 0.4\,{\rm s}$.\footnote{This is a simplified condition because the
reduced bias is a linear combination of the bias integrals for all parameters,
not just for $\cal M$.  However, as shown in Fig.~\ref{fig:bias integral}, the
bias integrals for different parameters have an approximately constant phase
difference.  If $|J_{\alpha}^{\rm net}|> \sqrt{3} |\bar{J}_\alpha|$ holds for
one parameter, it approximately holds for the others, as well as for the
quantities after linear combination.  Therefore, we simplify the condition by
comparing the reduced bias to the bias integral for a more concise
understanding.} Therefore, the network leads to larger biases than a single
detector for most $\Delta t_c$ values when only considering the time delay
effect.  Physically, this may imply that the spatial size of the network (around
the radius of the Earth) is not large enough to effectively separate  two
signals in time.  Though the conclusion sounds not intuitive since adding more
detectors is expected to give more information and helps PE, but it is important
to note that the reduced bias is defined as the ratio of the absolute bias and
the statistical uncertainty in Eq.~(\ref{eq:reduced bias}).  In a detector
network, the statistical uncertainty is reduced since the signals have a larger
SNR. Specifically in the case of Fig.~\ref{fig:J time delay}, for a large
absolute bias in the network, $|J_{\alpha}^{\rm net}|$ needs to be at least $3$
times of $|\bar{J}_\alpha|$, and no $\Delta t_c$ satisfies this condition.

We also simulate cases for the two signals placing at opposite ends of V1$-$H1
and V1$-$L1 lines, which correspond to the maximum arrival time difference
between the chosen detectors. We show $\Re J_{\alpha}$ as a function of $\Delta
t_c$ for these two cases in Fig.~\ref{fig:J time delay for three cases} along
with the above case of H1$-$L1 line.  In all three cases, $\Re J_{D\alpha}$
tends to cancel with each other for a small $\Delta t_c$, where the phase
difference is large. In the two new  cases, the maximum of $\Re J_{\alpha}^{\rm
net}$ is reached at $\sim 0.3\,{\rm s}$, and the maximum value also gives bias
integrals about thrice of that in a single detector, though the difference of
$\Delta \tau$ between the two detectors is even larger than that in the H1$-$L1
case ($\sim 50 \, {\rm ms}$ versus $\sim 20 \, {\rm ms}$).  In fact, around this
maximum point, the phase difference of $J_{D\alpha}$ between H1 (L1) and V1 is
about $2\pi$, which also leads to an aligned configuration. For a larger $\Delta
t_c$, the phase difference decreases, causing the three $J_{D\alpha}$ to first
become misaligned and then tend to realign.  Consequently, even in a network
capable of separating $J_{D\alpha}$ by several cycles, there still exists a
significant time span where the biases in the network are larger.  Recall that
the three cases shown in Fig.~\ref{fig:J time delay for three cases} correspond
to the maximum arrival time difference between the two detectors.  In realistic
scenarios, cases with a small difference are more common than those with a large
difference~\cite{Relton:2021cax}. For cases with a small $\Delta \tau$
difference, the phase difference between $J_{D\alpha}$ is also smaller, making
it even more likely for the network to have a larger bias than in a single
detector.  To investigate the overall impact of the time delay on the biases, in
Sec.~\ref{sec:average over extrinsic parameters} we present results averaged
over the sky location, which can be regarded as considering a realistic
distribution of $\Delta \tau$.

%---------------------------------------------------------------------
\subsection{Pattern function}\label{subsec:pattern function}
%---------------------------------------------------------------------

Considering different orientations of GW  detectors, the bias integral in the
$D$-th detector depends on the geometric parameters $\{\sky,\psi,\iota\}$
through $a_D$ and $\varphi_D$, where $a_De^{i\varphi_D} = F^D_+(1+\cos^2\iota)/2
- iF^D_\times\cos\iota$.  The bias integral in the $D$-th detector is given by
$J_{D\alpha} = \bar{J}_\alpha a_D^{(1)}a_D^{(2)}e^{i\Delta\varphi_D}/\bar{a}^2$.
Physically, this means to place all detectors at the Earth's center, and the
additional amplitude and phase come from combining two GW polarizations into the
strain in a detector with a specific orientation. For simplicity, we call this
the pattern function effect. 

When considering the time delay effect, the phase shift between $\bar{J}_\alpha$
and $J_{D\alpha}$ depends on the crossing frequency and decreases with $\Delta
t_c$. As a comparison, $a_D$ and $\varphi_D$ are constant for a given set of
geometric parameters.  The phase shift and amplitude ratio of $J_{D\alpha}$, as
well as their superposition $J_{\alpha}^{\rm net}$, from $\bar{J}_\alpha$, are
constant with respect to  $\Delta t_c$.  Similarly to Fig.~\ref{fig:J time
delay}, we show the bias integrals as a function of $\Delta t_c$ in
Fig.~\ref{fig:J pattern function}. We keep the same sky location of  two signals
in the H1$-$L1 line, and choose $\psi = 0$ and $\iota = 0$ for simplicity.
Differently from the case of time delay, in this case all bias integrals rotate
anticlockwisely with the same speed when $\Delta t_c$ increases, and the phase
difference between $J_{D\alpha}$ is constant.  This is also reflected in the
right panel of Fig.~\ref{fig:J pattern function}. The peaks of $\Re
J_{\alpha}^{\rm net}$ always appear at a same phase shift before (after) the
individual $J_{D\alpha}$, and the ratio of the peak height keeps constant.  We
also mark the maximum of $\Re J_{\alpha}^{\rm net}$ in the right panel (slightly
after $0.2\,{\rm s}$), and the corresponding $\Delta t_c$ approximately
maximizes $\bar{J}_\alpha$.  Due to the large phase difference between the three
$J_{D\alpha}$, the maximum of $\Re J_{\alpha}^{\rm net}$ is only about $1.4$
times of the maximum of $\Re J_{D\alpha}$, and the maximum of $|J_{\alpha}^{\rm
net}|$ is almost the same as that of $|\bar{J}_\alpha|$.  To convert the bias
integral to the reduced bias, note that the covariance matrix $\Sigma$ of the
SPE for the first signal in the network  is $\sum_{D=1} a_D^2/\bar{a}^2$ times
that in a single detector, which is approximately $2.3$ in our configuration. 
Therefore, $|J_{\alpha}^{\rm net}|$ needs to be approximately $\sqrt{2.3}
\approx 1.5$ times $|\bar{J}_\alpha|$ to ensure that the reduced bias in the
network exceeds that in a single detector.  This condition is not satisfied for
any $\Delta t_c$ in Fig.~\ref{fig:J pattern function}.

From the above discussion, it seems that the pattern function effect can
effectively reduce the biases in the network.  Indeed, the phase difference
between $J_{D\alpha}$ does not decrease with $\Delta t_c$, and it is intuitively
rare to find a case where the three $J_{D\alpha}$ are closely aligned. However,
when two signals are placed at opposite ends of the V1$-$H1 or V1$-$L1 lines,
the phase difference between $J_{D\alpha}$ is small enough to increase the
reduced bias in the network.  Apart from this, the biases in the network behave
similarly to the case of the H1$-$L1 line, so we do not show the corresponding
figures here for brevity.  To assess the influence of the pattern function on
the biases in a more comprehensive view, we need to average over the geometric
parameters $\{\sky,\psi,\iota\}$, which is discussed in Sec.~\ref{sec:average
over extrinsic parameters}.  Before ending this subsection, we emphasize that
the pattern function effect is independent of the size of the network, since
this effect arises from the different orientations of the detectors, regardless
of their locations. This is notably different from the time delay effect. 

%----
\begin{figure*}[t]
\centering
\includegraphics[width=0.8\textwidth]{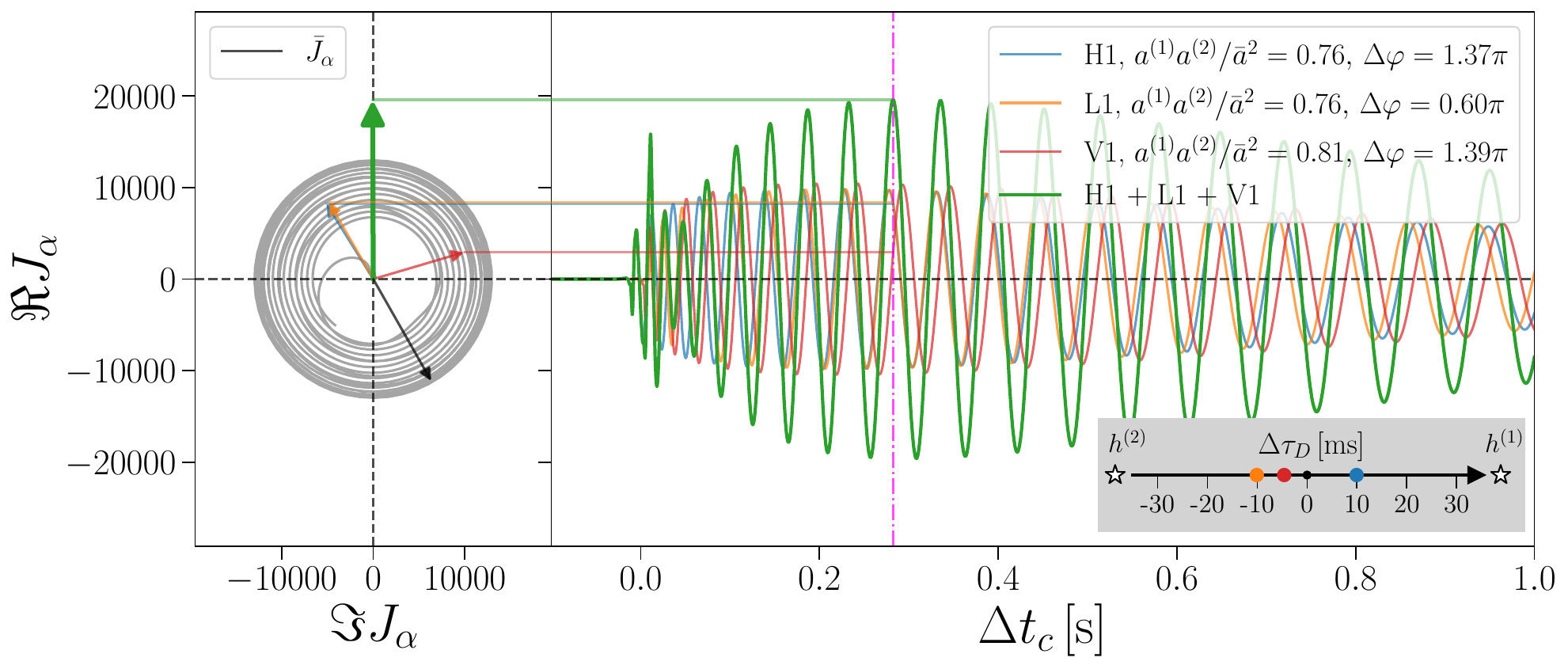}
\caption{Same as Fig.~\ref{fig:J pattern function}, but  including both the time
delay and the pattern function effects due to the different locations and
orientations of the detectors. The time delays, the additional amplitude and
phase factors (calculated from pattern functions) are marked. We also create an animated version
of this figure~\cite{supplemental_material}.}
\label{fig:J both effects}
\end{figure*}
%----

%---------------------------------------------------------------------
\subsection{Combined effects from time delay and pattern function}
\label{subsec:both effects}
%---------------------------------------------------------------------

Now we combine the two effects discussed in Sec.~\ref{subsec:time delay} and
Sec.~\ref{subsec:pattern function} together, and return to the bias integral in
Eq.~\eqref{eq:bias integral from J_bar}. Using the same parameter configuration
in plotting Fig.~\ref{fig:J pattern function}, Fig.~\ref{fig:J both effects}
shows the bias integrals with both features from the time delay and the pattern
function.  In this case, the bias integrals $J_{D\alpha}$ rotate anticlockwisely
with $\Delta t_c$ with a varying phase difference. The difference tends to a
constant, nonzero value for large $\Delta t_c$.  In the right panel of
Fig.~\ref{fig:J both effects}, one finds the destructive interference between
the three $\Re J_{D\alpha}$ for a small $\Delta t_c$ ($\lesssim 0.1\, {\rm s}$),
and the constant phase difference between them for a large $\Delta t_c$
($\gtrsim 0.6 \,{\rm s}$).  Mathematically, this is because that the phase shift
due to the time delay ($\sim -2\pi f^{\rm spa} \Delta \tau_D$) is larger for a
larger crossing frequency at a smaller $\Delta t_c$, thus dominates the
superposition for $\Delta t_c \lesssim 0.1 \,{\rm s}$. When comparing the network bias integral to the bias
integral in a single detector, the behavior resembles an ``average'' of the two
cases where only one effect is considered. The maximum value of $\Re
J_{\alpha}^{\rm net}$ lies between the maxima of $\Re J_{\alpha}^{\rm net}$ in
the time delay  and in the pattern function, and this maximum is also reached at
a $\Delta t_c$ between the maximum points of the two cases. 

In the simulated cases above, we artificially set  two sources at opposite ends
of the line connecting two detectors, which maximizes the time separation
between signals.  Although one still finds a time span where the biases in the
network are larger than that in a single detector, this configuration indeed
reduces the biases in the network to some extent compared to the worst case
where the two sources have the same extrinsic angle parameters.  In this case,
the two signals have the same time delay and pattern functions in all detectors,
and the impact of the network is only to increase the SNR and thus increase the
reduced biases.  In the network, all $J_{D\alpha}$ align with each other and
have the same evolution as $\bar{J}_\alpha$, which equivalently reduces to the
case in Ref.~\cite{Wang:2023ldq} with an angle-averaged waveform in a single
detector. 

%---------------------------------------------------------------------
\section{Population studies}
\label{sec:average over extrinsic parameters}
%---------------------------------------------------------------------

While Sec.~\ref{sec:case studies} provides a detailed discussion about how the
biases in different detectors are combined in the network with their dependence
on the coalescence time difference $\Delta t_c$, it is based on some specific
illustrative configurations of the extrinsic angle parameters
$\big\{\sky,\psi,\iota\big\}$.  To highlight the effect of the network on the
biases, these configurations are artificially designed to maximize the time
delay between two signals in different detectors. 
% However, there can also be cases with similar time delays and pattern functions
% of the two signals in different detectors, which are even more common in
% realistic scenarios~\cite{Relton:2021cax}. 
To draw a conclusion about the general influence of the network on the biases,
in this section we conduct a population-level simulation to average over the
extrinsic parameters.

We first describe the simulation setup. We choose the same intrinsic parameters
as in Sec.~\ref{subsec:bias in complex plane}, with  two signals respectively
having $m^{(1)}_1 = 30\,\mathrm{M_\odot}$, $m^{(1)}_2 = 20\,\mathrm{M_\odot}$,
and $m^{(2)}_1 = 27\,\mathrm{M_\odot}$, $m^{(2)}_2 = 18\,\mathrm{M_\odot}$ in
the detector frame.  We also fix the luminosity distance to $d_L^{(1)} =
10\,\mathrm{Gpc}$ and $d_L^{(2)} = 60\,\mathrm{Gpc}$.  Here we are more
interested in comparing the biases in the network and in a single detector, and
changing $d_L$ only gives an overall scaling for all biases. We sample the
remaining six extrinsic parameters: the geometric angle parameters
$\{\sky,\psi,\iota\}$, and the coalescence time $t_c$ and phase $\phi_c$.  We
assume that the sources have isotropic sky locations and orbital orientations
with uniformly distributed coalescence phase $\phi_c$, that is, drawing samples
with $\alpha \sim {\cal U}(0, 2\pi)$, $\sin \delta \sim {\cal U}(-1, 1)$,  $\cos
\iota \sim {\cal U}(-1, 1)$, $\psi \sim {\cal U}(0, \pi)$, and $\phi_c \sim
{\cal U}(0, 2\pi)$.  For the  time, we directly sample the difference
$\Delta t_c = t_c^{(2)} - t_c^{(1)}$ from a uniform distribution ${\cal U}(0\,
{\rm s},5\, {\rm s})$.  We only consider the positive $\Delta t_c$ since the
second source is lighter and the time-frequency crossing exists only when the
second source merges later~\cite{Wang:2023ldq}.  The upper limit of $5\,{\rm s}$
is a conservative choice, slightly larger than the threshold criterion for
overlapping signals, as revealed in previous studies~\cite{Himemoto:2021ukb,
Pizzati:2021apa, Hu:2022bji}. The dependence of the time threshold is discussed
later.  Adopting more realistic distributions for $\Delta t_c$, such as the
exponential distribution~\cite{Relton:2021cax} or the scaled beta
distribution~\cite{Wang:2025ckw}, does not affect the results significantly.  We
still use the 3-detector network with the CE-2 sensitivity as described in
Sec.~\ref{sec:case studies}. The number of samples is sufficient to
ignore the statistical fluctuations.

For each sample, we calculate the bias integral $J_{\alpha}^{\rm net}$ in the
network and the reduced bias $B_\alpha^{\rm net}$ with the corresponding
covariance matrix $\Sigma$. The variable parameters are chosen as $\big \{ {\cal
M}, \eta, d_L, t_c \big\}$.  As discussed in Sec.~\ref{subsec:bias in complex
plane}, the modulus of the bias integral, as well as
$|\Sigma^{\alpha\beta}J_\beta|$ after being combined with the covariance matrix,
serves as a better measure for the bias.  Therefore, we also record
$|J_{\alpha}^{\rm net}|$ and $\max_{\Delta \phi_c} \left|B_\alpha^{\rm net} \right|$ for each
sample. Similarly to Sec.~\ref{sec:case studies}, we choose three scenarios to
compare the biases in the network and in a single detector: (i) only considering
the time delay effect; (ii) only considering the pattern function effect; and
(iii) considering both effects. \Referee{We calculate corresponding quantities
in a single detector ignoring geometric effects, which, given the mass
parameters and the luminosity distance, only depend on $t_c$ and $\phi_c$ of the
samples.}

%----
\begin{figure}[t]
\centering
\includegraphics[width=0.48\textwidth]{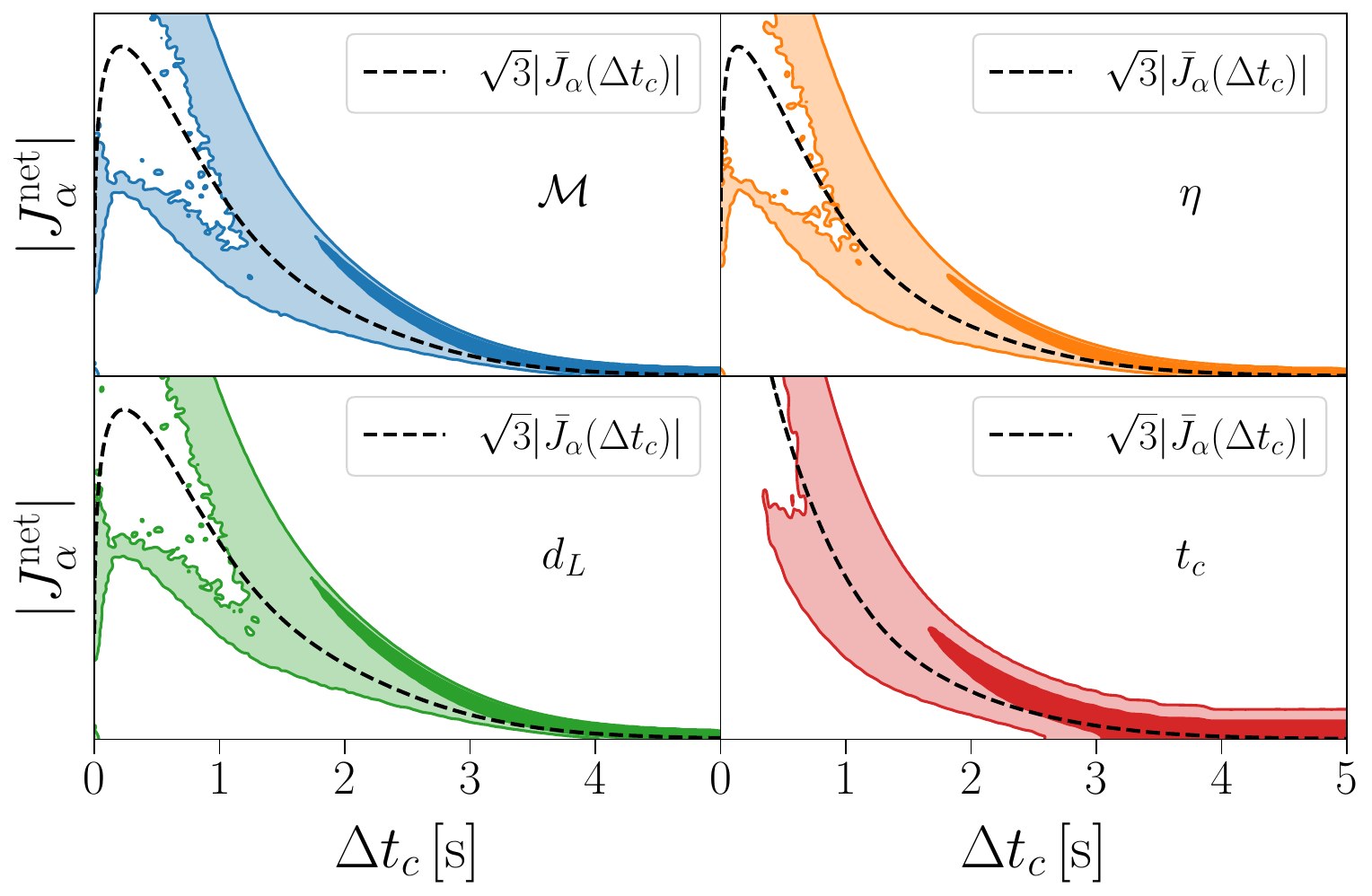}
\caption{Joint distribution between the modulus of the bias integral
$|J_{\alpha}^{\rm net}|$ and the coalescence time difference $\Delta t_c$ for
four variables $\cal M$, $\eta$, $d_L$, and $t_c$. In the simulation we only
consider the time delay effect. Contours show the 50\% and 90\% levels of the
distribution. The black dashed line indicates the threshold
$\sqrt{3}|\bar{J}_\alpha(\Delta t_c)|$, above which the points are expected to
have a larger reduced bias in the network than in a single detector. }
\label{fig:Js distribution}
\end{figure}
%----

%----
\begin{figure*}[t]
\centering
\includegraphics[width=0.6\textwidth]{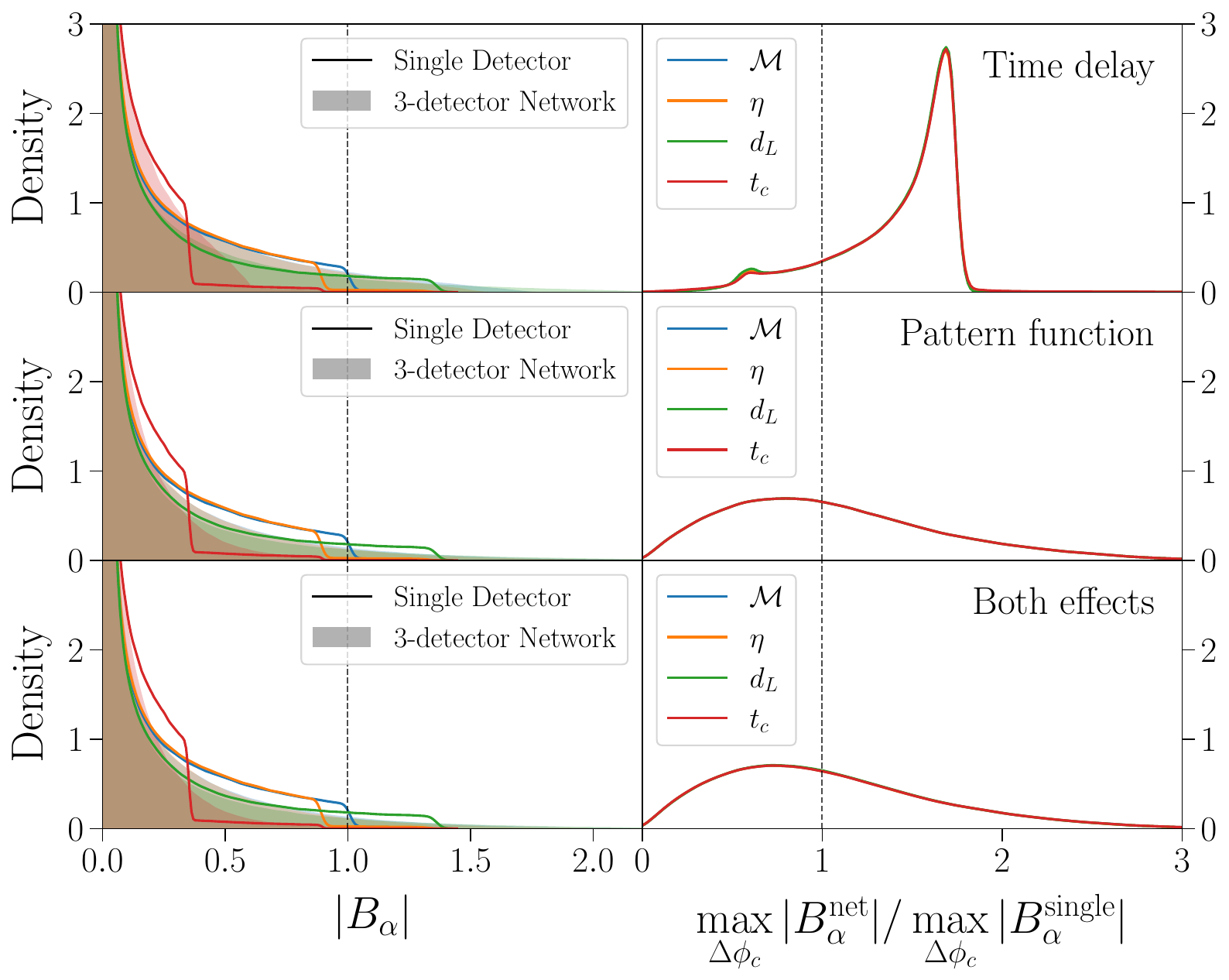}
\caption{Distribution of $|B_\alpha|$ in the network and in a single detector
(left), and the ratio of the maximum reduced bias $\max_{\Delta \phi_c} \left|
B_\alpha \right|$ between the network and a single detector (right). Three rows
correspond to the three scenarios (from top to  bottom): (i) only considering
the time delay effect, (ii) only considering the pattern function effect, and
(iii) considering both effects.  In the left column, black dashed lines indicate
$|B_\alpha| = 1$, exceeding which means that the overlapping signals lead to a
significant bias compared to the statistical uncertainty. In the right column,
black dashed lines indicate the ratio of 1, exceeding which means that the
network worsens the influence of the overlapping and leads to a larger reduced
bias than in a single detector.}
\label{fig:compare bias}
\end{figure*}
%----

Before comparing the results in a single detector and in the network, we first
demonstrate the relationship between the case studies and the population-level
simulation. Taking the scenario of including only the time delay effect as an
example, we show the joint distribution of $|J_{\alpha}^{\rm net}|$ and $\Delta
t_c$ in Fig.~\ref{fig:Js distribution}.  We also plot the line of
$\sqrt{3}|\bar{J}_\alpha(\Delta t_c)|$ in the figure, which represents the
threshold for having a larger reduced bias in the network than in a single
detector, as explained in Sec.~\ref{subsec:time delay}.  Generally, the four
variables have similar joint distributions, and the contours approximately have
the same shape  trending with the $\sqrt{3}|\bar{J}_\alpha(\Delta t_c)|$ line
since $J_{\alpha}^{\rm net}$ is just the superposition of the bias integrals in
different detectors.  We find more than half of the samples fall above the
$\sqrt{3}|\bar{J}_\alpha(\Delta t_c)|$ line, indicating that the network
increases the reduced biases for a large portion of  samples.  More notably, 
samples above the line are more likely to have a larger $\Delta t_c$, which is
consistent with the case studies in Sec.~\ref{subsec:time delay}.  For a large
$\Delta t_c$, the crossing frequency of  two signals decreases, making
$J_{D\alpha}$ in different detectors more aligned and resulting in a larger
$|J_{\alpha}^{\rm net}|$.  Numerically, in our configuration, $\Delta t_c
=3\,{\rm s}$ approximately corresponds to a crossing frequency of $8\,{\rm Hz}$.
When combined with the typical time delay (the mean of $\Delta \tau$ is around
$10$--$20\,{\rm ms}$), the phase difference between $J_{D\alpha}$ is around
$0.2\pi$, which is small enough to align the signals.

%----
\begin{table}[t]
    \renewcommand{\arraystretch}{1.5} % 调整行间距
    \setlength{\tabcolsep}{0.25cm} % 调整列间距
    \centering
    \caption{Fraction of samples with the magnitude of the reduced bias
    $|B_\alpha|$ larger than 1.  Different rows correspond to the
    single-detector scenario and the three network scenarios with different
    effects included. All statistics are rounded to two significant digits.}
    \vspace{0.cm}\label{tab:fraction of larger bias parameters}
    \begin{tabular}{lcccc}
    \hline\hline
    Parameter & $\cal M$ & $\eta$ & $d_L$ & $t_c$\\
    \hline
    Single detector & 0.63\% & 0.65\% & 5.7\% & 0.016\% \\
    \hline
    Time delay & 8.2\% & 6.2\% & 8.1\% & 0.38\% \\
    Pattern function & 6.0\% & 4.8\% & 6.8\% & 0.58\% \\
    Both effects & 5.4\% & 4.2\% & 6.2\% & 0.41\% \\
    \hline
    \end{tabular}
\end{table}
%----

%----
\begin{table}[t]
    \renewcommand{\arraystretch}{1.5} % 调整行间距
    \setlength{\tabcolsep}{0.3cm} % 调整列间距
    \centering
    \caption{Fraction of samples whose biases are larger in the network than in
    a single detector. The difference between the four parameters is negligible,
    and only the average over them is shown.  Different rows correspond to the
    three scenarios with different effects included in the network. All
    statistics are rounded to two significant digits.}
    \vspace{0.cm}
    \label{tab:fraction of B and maxB}
    \begin{tabular}{lcc}
    \hline\hline
     & $|B_{\alpha}|$ & ${\max}_{\Delta \phi_c} \left|B_{\alpha} \right|$ \\
    \hline
    Time delay & 69\% & 87\% \\
    Pattern function & 47\% & 49\% \\
    Both effects & 46\% & 47\% \\
    \hline
    \end{tabular}
\end{table}
%----

We show the results of our simulations in Fig.~\ref{fig:compare bias}, where the
three rows correspond to the three scenarios. In the left column, we plot the
distribution of $|B_\alpha|$ in the network and in a single detector, and the
right column shows the ratio of the $\max_{\Delta \phi_c} \left| B_\alpha
\right| $ between the network and a single detector.  We also summarize some
statistics of the samples in Table~\ref{tab:fraction of larger bias parameters}
and Table~\ref{tab:fraction of B and maxB}.  For $|B_\alpha|$, we find that in
all three scenarios, the network increases biases overall, extending the
distribution to larger values. This is also evident from the significant
difference in the fraction of samples with $|B_\alpha| > 1$ between the network
and a single detector, as shown in Table~\ref{tab:fraction of larger bias
parameters}.  The difference in the distribution among the four parameters
arises from their distinct contributions to the waveform, i.e., the ${\cal
A}_{,\alpha}-i\phi_{,\alpha}{\cal A}$ term in Eq.~\eqref{eq:J_bar}, which leads
to varying magnitudes of  biases. In fact, the absolute magnitude of biases also
depends on  SNRs of two signals. Comparing $|B_\alpha|$ with 1 intuitively
represents the influence of overlapping but may not be an effective way to
characterize the contributions from the network. However, if we compare the
biases between the network and a single detector, all four parameters have
almost the same results as shown in the right column of Fig.~\ref{fig:compare
bias}.  The difference of the fraction of samples with a larger $|B_\alpha|$
(and $\max_{\Delta \phi_c} \left| B_\alpha \right| $) in the network among the
four parameters is less than $1\%$, and in Table~\ref{tab:fraction of larger
bias parameters} we only show their average over the four parameters.  From the
right column of Fig.~\ref{fig:compare bias} and Table~\ref{tab:fraction of B and
maxB}, it is  clear that a large fraction of samples have larger biases in the
network than in a single detector for all three scenarios.

%----
\begin{figure*}[t]
\centering
\includegraphics[width=0.7\textwidth]{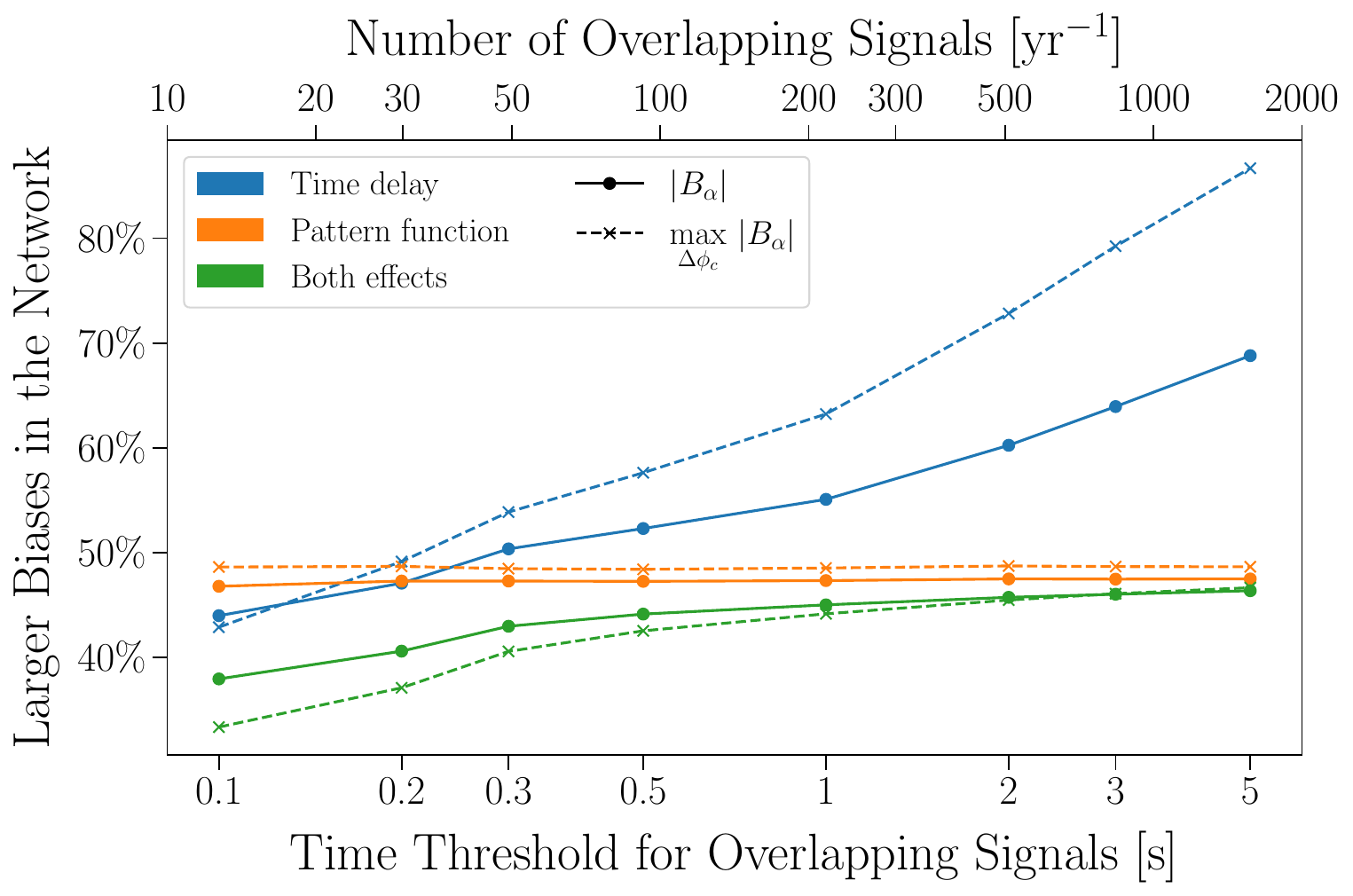}
\caption{Fractions of samples with larger $|B_\alpha|$ and $\max_{\Delta \phi_c}
\left| B_\alpha \right|$ in the network than in a single detector, as functions
of the time threshold for overlapping signals. Blue, orange, and green colors
correspond to the scenarios with (i) only considering the time delay effect,
(ii) only considering the pattern function effect, and (iii) considering both
effects, respectively. The solid lines and the dots represent for $|B_\alpha|$,
while the dashed lines and the crosses represent for $\max_{\Delta \phi_c}
\left| B_\alpha \right| $.  The expected numbers of overlapping events for
different thresholds are calculated from corresponding
formula~\cite{Wang:2025ckw} with an assumed detection rate of $10^5\,{\rm
yr}^{-1}$~\cite{Pizzati:2021apa, Samajdar:2021egv, Himemoto:2021ukb,
Hu:2022bji}.
}\label{fig:fraction of larger bias}
\end{figure*}
%----

Now we go deeper into the three scenarios in the right column of
Fig.~\ref{fig:compare bias}. For the scenario only considering time delay, most
samples exhibit a larger $\max_{\Delta \phi_c}|B_\alpha|$ in the network,
consistent with the case studies in Sec.~\ref{subsec:time delay} and
Fig.~\ref{fig:Js distribution}.  When $\Delta t_c \gtrsim 1\,{\rm s}$, the delay
from the Earth center to the detectors is insufficient to effectively separate
the signals. More detectors merely increase their SNR and result in larger bias.
For a large $\Delta t_c$, the $J_{D\alpha}$ tend to align, making
$J_{\alpha}^{\rm net}$ roughly three times of $\bar{J}_\alpha$.  Besides, the
network's covariance matrix $\Sigma$ is one third of that in a single detector.
Therefore, the samples with a large $\Delta t_c$ approximately have a $\sqrt{3}$
times larger $\max_{\Delta \phi_c}|B_\alpha|$ in the network, explaining the
peak around 1.7 in Fig.~\ref{fig:compare bias}.  We also find a small difference
between the distribution for different parameters. This is because that the
evolution of $\bar{J}_\alpha$ is only approximately the same for different
parameters (the first term in the square bracket of Eq.~\eqref{eq:derivative of
J_bar}).  When shifting the variable $\Delta t_c$ to $\Delta t_c+\Delta \tau$,
all $\bar{J}_\alpha$ do not strictly change in the same way, leading to a small
difference in the final results.  For the scenario only considering  pattern
function, the ratio distribution is smoother, extending to larger values.  This
is because that the amplitude term $a^{(1)} a^{(2)}/\bar{a}^2$ has a wider range
of values, and the phase shift $\Delta \varphi$ is approximately uniformly
distributed in $[0, 2\pi]$ for  randomly sampled locations and orientations of
GW sources.  Also, since the pattern function effect is equivalent to
multiplying $\bar{J}_\alpha$ by a constant complex number which is independent
of parameters, the increase/decrease of $J_{\alpha}^{\rm net}$ is strictly the
same for all parameters.  In the second row of the right column of
Fig.~\ref{fig:compare bias}, all four parameters have the same distribution. For
the scenario with both effects, the distribution is very similar to the scenario
only considering pattern function, with almost invisible differences for
different parameters.  As estimated above, the phase shift due to the time delay
is around $0.2\pi$ for most samples, while the phase shift due to the pattern
function is approximately uniformly distributed in $[0, 2\pi]$. Therefore, when
combining  two effects, the pattern function effect dominates the superposition
of the bias integrals, leading to a similar distribution as for the case only
considering  pattern function.  This is also shown in the case studies in
Sec.~\ref{subsec:both effects}, where the bias integral behaves like the case
only considering time delay for a small $\Delta t_c$ and like the case only
considering pattern function for a large $\Delta t_c$.

%----
\begin{figure*}[t]
\centering
\includegraphics[width=0.75\textwidth]{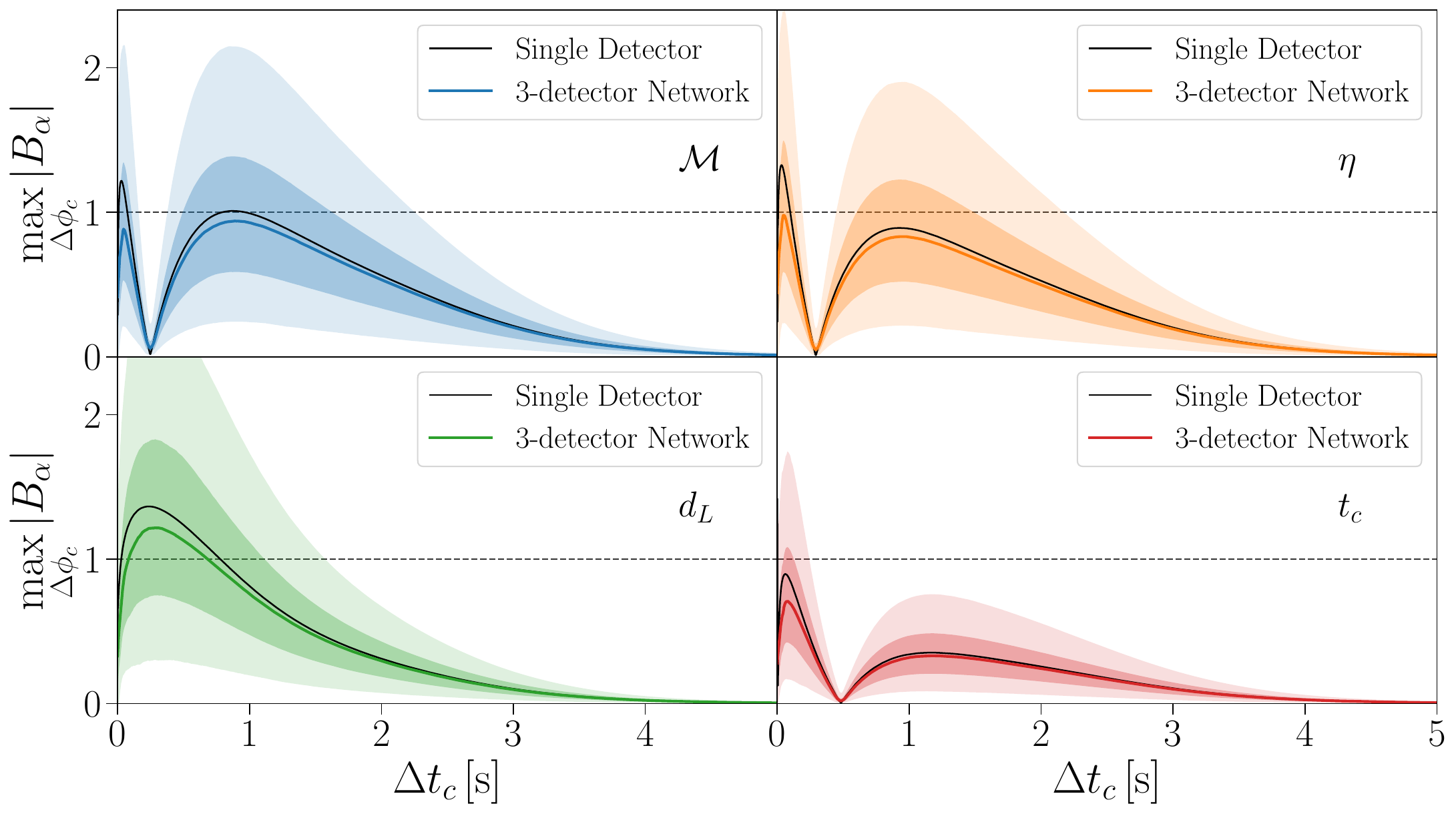}
\caption{Relationship between the maximum reduced bias 
$\max_{\Delta\phi_c}\left| B_\alpha \right|$ and the coalescence-time difference
$\Delta t_c$ for parameters $\mathcal{M}$, $\eta$, $d_L$, and $t_c$.  The black
curve shows the single-detector result ignoring the time delay and pattern
function effects.  For the network case, both the time-delay and
pattern-function effects are included, and we plot the median together with the
$50\%$ and $90\%$ central intervals of the simulated samples for each $\Delta
t_c$.  The black dashed line marks the threshold of $1$, above which the bias
becomes significant compared with the statistical uncertainty.
}\label{fig:bias distribution network}
\end{figure*}
%----

Next, we discuss how the influence of network on the biases varies with the
time threshold for overlapping signals. Determining an appropriate criterion to 
indicate overlapping signals is still an open question, and a simple criterion
is establishing a time threshold $\Delta t_{\rm th}$, which is estimated to be
ranging from ${\cal O}(0.1\,{\rm s})$ to ${\cal O}(1\,{\rm
s})$~\cite{Relton:2021cax, Himemoto:2021ukb, Pizzati:2021apa, Antonelli:2021vwg,
Samajdar:2021egv}. Different criteria lead to different estimation for the
number of overlapping events in the future observations, affecting the search
and PE strategies in  data analysis.  Here we do not aim to discuss the optimal
criterion or the strategy for analyzing overlapping signals, but discuss how a
network influences the biases for different thresholds $\Delta t_{\rm th}$,
corresponding to overlapping signals with different time separation. In
Fig.~\ref{fig:fraction of larger bias}, we plot  the fractions of samples with
larger $|B_{\alpha}|$ and $\max_{\Delta \phi_c} \left| B_{\alpha} \right|$ in
the network than in a single detector (averaged over the four parameters), as
functions of $\Delta t_{\rm th}$. Adopting a detection rate of $10^5$ BBH
mergers per year~\cite{Pizzati:2021apa, Samajdar:2021egv, Himemoto:2021ukb,
Hu:2022bji} and the formalism for calculating the number of overlapping
events~\cite{Wang:2025ckw}, we  mark the expected number of overlapping events
per year for different $\Delta t_{\rm th}$. 

In the scenario with only the time delay effect, the fractions of samples with
larger biases increase with $\Delta t_{\rm th}$. As explained above, overlapping
signals with a large $\Delta t_c$ are more likely to have larger biases in the
network. On the other hand, the phase shift due to the pattern function effect
is independent of $\Delta t_c$. Therefore, in the scenario with only the pattern
function effect, the fractions are constant with respect to $\Delta t_{\rm th}$.
When combining the two effects, the fractions slowly increase with $\Delta
t_{\rm th}$, and tend to the constant value in the case only considering pattern
function.  This is because that the phase shift due to the time delay is small
for large $\Delta t_c$. Though this helps to align  $J_{D\alpha}$ when only
considering the time delay effect, it fails to dominate the superposition when
the pattern function effect with $\Delta \varphi\sim {\cal O}(\pi)$ is present. 
In addition, the fractions when combining the two effects are always smaller
than those when only one effect is considered, which is intuitively expected
since the inclusion of both effects leads to more differences between the bias
integrals in different detectors. \Referee{Notably, the fractions of samples with larger biases in the network are about
$40\%$ or higher for all $\Delta t_{\rm th}\in [0.1, 5]\,{\rm s}$, even both
geometric effects are considered.}

\Referee{Though we have found that a significant portion of the samples exhibit
larger biases in the network than in a single detector, it remains unclear
whether these biases are more significant than the statistical uncertainties. 
In Fig.~\ref{fig:bias distribution network}, we plot $\max_{\Delta\phi_c}\left| B_\alpha \right|$ as a function
of $\Delta t_c$ for both the single-detector scenario and the network scenario
with geometric effects included.  For the network case, we plot the distribution
of $\max_{\Delta\phi_c}\left| B_\alpha \right|$ with the median, the $50\%$ and
$90\%$ central intervals at each fixed $\Delta t_c$, since the biases also
depend on the simulated extrinsic parameters.  Taking the chirp mass $\cal{M}$
as an example, in the single-detector scenario the bias exceeds $1$ only within
a small region with $\Delta t_c \lesssim 0.1\,{\mathrm s}$, and the height of
the second peak around $\Delta t_c \sim 1\,{\mathrm s}$ is close to $1$.  In the
network scenario, the distribution of $\max_{\Delta\phi_c}\left| B_{\alpha}
\right|$ is roughly centered around the single-detector curve due to the
different combinations of extrinsic parameters, and nearly half of the samples
yield biases larger than the single-detector values.  Consequently, there are
more samples with $\max_{\Delta\phi_c}\left| B_{\alpha} \right| > 1$ over a
broad range of $\Delta t_c$, especially near the second peak. The profile of the
distribution at fixed $\Delta t_c$ in fact provides another perspective on the
lower right panel of Fig.~8, which similarly shows that nearly half of the
samples satisfy $\max_{\Delta\phi_c}\left| B_{\alpha}^{\mathrm{net}} \right| >
\max_{\Delta\phi_c}\left| B_{\alpha}^{\mathrm{single}} \right|$. For a large
$\Delta t_c$ (e.g., $\gtrsim 3\,{\mathrm{s}}$), although the network still
increases the biases for a substantial fraction of samples, the absolute
magnitudes of the biases become negligible compared with the statistical
uncertainties, consistent with the intuition that for a large $\Delta t_c$ the
two signals are well separated in time and the effect of overlapping is limited.
In addition, if the amplitude of the second signal is changed, the magnitude of
the biases will be scaled accordingly, but the ratio between the biases in the
network and in a single detector still remains unchanged.}

\Referee{The findings in Fig.~\ref{fig:bias distribution network} demonstrate that, although for a
large time separation the absolute magnitudes of the biases may be negligible
compared with the statistical uncertainties, the network still extends the range of
$\Delta t_c$ where the overlapping signals lead to significant biases. Therefore, within a detector network, the
overlapping signals still significantly affect data analysis.}

%---------------------------------------------------------------------
\section{Summary and Discussion}\label{sec:summary and discussion}
%---------------------------------------------------------------------

In the data stream of XG GW detectors, such as CE and ET, overlapping signals
are expected to arise due to the increased detection rate and signal duration,
posing challenges to  data analysis. If not handled properly, the overlapping
signals can cause substantial biases in PE, possibly misleading subsequent
astrophysical conclusions.  In developing strategies for addressing overlapping
signals, it is helpful to investigate how the biases form and what is their
dependence on the signal parameters. In this work, we extend the previous study
of the bias behavior over intrinsic parameters in a single
detector~\cite{Samajdar:2021egv, Pizzati:2021apa, Antonelli:2021vwg, Hu:2022bji,
Wang:2023ldq, Dang:2023xkj} to a detector network, considering the geometric
effects due to the different locations and orientations of the detectors. We
conduct a quantitative anatomy about how the biases depend on the extrinsic
parameters, including the sky location, polarization and inclination angles, as
well as the coalescence time and phase, and compare the biases in a network with
those in a single detector. 

Similarly to the formalism in \citet{Wang:2023ldq}, we analyze the contributions
to the biases from geometric effects by explicitly showing the dependence on the
parameters in Sec.~\ref{sec:analytical expression}.  By introducing the bias
integral $J$, we can directly separate the geometric effects from the intrinsic
parameters.  As shown in Eq.~\eqref{eq:bias integral from J_bar} and
Fig.~\ref{fig:illustration of bias integral}, the bias integral in any detector
in the network is generated from the bias integral $\bar{J}$ in a fictitious
detector located at the Earth center with angle-averaged response, while the
relationship between them only depends on the angle parameters
$\big\{\sky,\psi,\iota\big\}$ and the detector's location and orientation.  The
bias integral in the network is simply the superposition of those in individual
detectors, and the increase/decrease of the biases in the network roughly
depends on the degree of alignment of the individual bias integrals in the
complex plane.

With the help of SPA, we are able to theoretically study the evolution of the
bias integral with the coalescence time difference $\Delta t_c$ between the
overlapping signals. In the complex plane, the bias integral in a single
detector rotates as $\Delta t_c$ increases with a slowly varying amplitude,
explaining the oscillatory behavior of the biases observed in previous
studies~\cite{Pizzati:2021apa, Himemoto:2021ukb, Wang:2023ldq}.  In the network,
the integrals in different detectors co-rotate with $\bar{J}$ while interfering
with each other, resulting in more complicated behavior of the biases. In
Sec.~\ref{sec:case studies}, we decompose the geometric effects into two parts 
and discuss them separately, where the location effect introduces additional
time delays while the orientation effect introduces additional amplitude and
phase factors. The location effect equivalently shifts the variables $\Delta
t_c$ in the bias integrals in different detectors, and becomes less significant
for a large $\Delta t_c$. As a comparison, the additional factors from the
orientation effect are independent of $\Delta t_c$. 

For a more systematic and comprehensive assessment for the potential influence
of the network on the biases, in Sec.~\ref{sec:average over extrinsic
parameters} we simulate a population of overlapping signals with varying
extrinsic parameters as an extension of the case studies in Sec.~\ref{sec:case
studies} and previous work~\cite{Relton:2021cax}. Nearly half of the samples
have larger biases in the network than in a single detector, even when both
geometric effects are considered. This indicates that overlapping signals still
significantly affect  PE in a network, highlighting the need for a careful
consideration when designing strategies for analyzing overlapping signals. A
network increases the SNR of the signals, thus reducing the statistical
uncertainty and making the biases more significant. In addition, this implies
that the spatial size of the network, around the scale of the Earth, is not
large enough to effectively separate the overlapping signals in time. For
example, if the light travel time between the detectors is larger than the decay
timescale of the bias integral, only one detector in the network will have a
substantially non-zero $J$, after combining with the covariance matrix, the
biases are roughly $1/\sqrt{N_D}$ times of that in a single detector, with the
number of detectors $N_D$. 

There are several possible extensions for studying the biases due to overlapping
signals in the XG detector network.  First, throughout the paper we fixed the
masses of  two signals as one representative case. Similar to the
source-dependent threshold for overlapping signals~\cite{Relton:2021cax,
Johnson:2024foj}, sources with different masses have different decay timescales
of the bias integral. Though this does not affect the qualitative conclusions in
this work, including the distribution of intrinsic parameters in the simulation
makes the fractions in Table~\ref{tab:fraction of B and maxB} more practical.
\Referee{Second, in this work we focus on how extrinsic angle parameters
$\big\{\sky,\psi,\iota\big\}$ affect PE biases of $\big\{ {\cal M}, \eta, d_L,
t_c \big\}$, while the angle parameters themselves can also be biased. Relaxing
these angle parameters in the PE does not significantly change the bias behavior
of the intrinsic parameters (${\cal M}$ and $\eta$) due to their weak
correlation, but the biases in $d_L$ and $t_c$ can be affected. The relationship between the bias integrals of the angle parameters 
 in different detectors also becomes more
complicated, calling for a more comprehensive study.} 
 The
overlapping signals may result in fake precession
features~\cite{Relton:2021cax}, suggesting a need to study biases in spin
parameters. In addition, we have ignored the Earth's rotation and the finite
size of detectors, so the geometric effects are constant during the signal
duration. This assumption approximately holds for BBH signals lasting from
minutes to hours, but breaks down for signals from binary neutron stars that
last from hours to days~\cite{Himemoto:2021ukb, Wu:2022pyg, Johnson:2024foj}.
Including  frequency-dependent pattern functions~\cite{Essick:2017wyl,
Chen:2024kdc} may introduce new behaviors of  biases. Correlated noise of
detectors can also have impacts~\cite{Cireddu:2023ssf, Wong:2024hes}. In
addition, there are other sources of biases in the PE, such as from the waveform
modelling~\cite{Hu:2022bji} and glitches~\cite{Ghonge:2023ksb}.  We leave these
investigations to future work. 

\Referee{Another important aspect is that here we only consider the biases in SPE, where
only one template is used to fit the overlapping signals while ignoring the
existence of the other signal. When the second signal is below the detection
threshold, our results directly reflect the influence of overlapping on
analyzing the first signal (the solely detectable one). 
When both signals are loud enough to be detectable, it is more appropriate to adopt
the hierarchical-subtraction
strategy~\cite{Antonelli:2021vwg, Janquart:2022fzz, Hu:2025vlp} or conduct a
joint PE~\cite{Janquart:2022fzz, Baka:2025yqx}. SPE can be regarded as the first
step of hierarchical subtraction, and our analysis shows in a detector network that
this subtraction step may be more biased than anticipated. This
motivates a comparison of the final results from hierarchical subtraction in a
network and in a single detector, which we leave for future work.}

Also, though the existence
of overlapping signals and the magnitude of biases
cannot be determined as a priori, if there are some ways
to identify them, it is better to conduct a joint PE to avoid large biases. In this
strategy, the SPE biases are still useful to quantify the influence of
overlapping. For example, \citet{Baka:2025yqx} compared the SPE posteriors with
the original data and the data subtracted by the best-fit waveform of the other
signal. If the two posteriors are significantly different, it indicates that the
influence of overlapping and the signals should be analyzed jointly. Based on
the bias behaviors studied in this work, we propose another possible strategy
for identifying overlapping signals from a network perspective. As shown in
Fig.~\ref{fig:J pattern function} and Fig.~\ref{fig:J both effects}, there are
large phase differences between the bias integrals in different detectors.
\Referee{Therefore, if one conducts SPE in different detectors or sub-networks, the
posteriors may have large differences, and this inconsistency has the potential to serve as a criterion for identifying overlapping signals.} In addition, in
a detector network, it is possible to construct the so-called null
stream~\cite{Guersel:1989th, Chatterji:2006nh} by linearly combining data from
multiple detectors to eliminate the GW signals from a specific sky location.
\Referee{
This method was first proposed for GW detections~\cite{Guersel:1989th,
Chatterji:2006nh, Chu:2020pjv} and then was applied to search
for extra polarization modes beyond GR~~\cite{LIGOScientific:2020tif,LIGOScientific:2021sio,Hagihara:2018azu,Hagihara:2019rny,Pang:2020pfz,Wong:2021cmp,Zhang:2021fha,Hu:2023soi,Liang:2024sfn}, while also has the
potential to search for subordinate overlapping signals in the data.} The bias behaviors
studied in this work shall provide  insights for developing methods to analyze
overlapping GW signals in the future GW detector network.

%---------------------------------------------------------------------
\begin{acknowledgments}
\Referee{We thank the anonymous referee for the valuable comments.}
This work is supported by the National Natural Science Foundation of China
(123B2043,  12573042, 12405065, 12465013), the National SKA Program of China
(2020SKA0120300), the Beijing Natural Science Foundation (1242018), the Max
Planck Partner Group Program funded by the Max Planck Society, and the
High-performance Computing Platform of Peking University. 
\end{acknowledgments}
%---------------------------------------------------------------------

\bibliography{network_refs.bib}{}% Produces the bibliography via BibTeX.

\end{document}